\documentclass[superscriptaddress, twocolumn, prd, showpacs, nofootinbib, a4]{revtex4-1}
\usepackage{graphicx}
\usepackage{enumerate}
\usepackage{tikz}
\usepackage{amsmath}
\usepackage{amssymb}
\usepackage{commath}
\usepackage{bbold}
\usepackage{hyperref} 
\usepackage{mathtools}
\usepackage{listings}
\usepackage{xcolor}
\usepackage{xspace}

\usepackage{dsfont}
\usepackage{hyperref}
\usetikzlibrary{arrows,positioning}

\newcommand{\swordfish}{\textsf{swordfish}\xspace}
\newcommand{\vect}[1]{\boldsymbol{\mathbf{#1}}}

\newcommand{\TS}{\text{TS}}
\newcommand{\nbins}{{n_\text{b}}}

\begin{document}

\title{\textsf{swordfish:} Efficient Forecasting of New Physics Searches
without Monte Carlo}

\author{Thomas D. P. Edwards}
\email{t.d.p.edwards@uva.nl}
\author{Christoph Weniger}
\email{c.weniger@uva.nl}

\affiliation{Gravitation Astroparticle Physics Amsterdam (GRAPPA), Institute
for Theoretical Physics Amsterdam and Delta Institute for Theoretical Physics,
University of Amsterdam, Science Park 904, 1090 GL Amsterdam, The Netherlands}

\date{Compiled on \today}

\begin{abstract}
  We introduce \swordfish,\footnote{\href{http://www.github.com/cweniger/swordfish}{github.com/cweniger/swordfish}} a
  Monte-Carlo-free \textsf{Python} package to predict expected exclusion limits,
  the discovery reach and expected confidence contours for a large class of
  experiments relevant for particle- and astrophysics.  The tool is applicable
  to any counting experiment, supports general correlated background
  uncertainties, and gives exact results in both the signal- and
  systematics-limited regimes.  Instead of time-intensive Monte Carlo
  simulations and likelihood maximization, it internally utilizes new
  approximation methods that are built on information geometry.  Out
  of the box, \swordfish\ provides straightforward methods for accurately
  deriving many of the common sensitivity measures.  
  In addition, it allows one to examine experimental abilities in great detail by
  employing the notion of \emph{information flux}. This new concept generalizes
  signal-to-noise ratios to situations where background uncertainties and
  component mixing cannot be neglected.
  The user interface of \swordfish\ is designed with ease-of-use in mind, which
  we demonstrate by providing typical examples from indirect and direct dark
  matter searches as \textsf{jupyter notebooks}.
\end{abstract}

\maketitle

\section{Introduction}

Sensitivity to signs of new physics is the most desirable feature of a future
experiment. As such, optimizing the process by which these sensitivities are
calculated is of utmost importance for efficient scientific progress.
Calculating the projected sensitivity of various experiments has proven, in the
past, to be a computationally expensive task. From experimentalists looking to
optimize the parameter space they rule out to theorists calculating whether
their particle physics model will be observable in the future, sensitivity
projections are a ubiquitous task among physicists and astronomers.

The most commonly used sensitivity calculations are based on the maximum
likelihood ratio (MLR) method which, in the large-sample (Gaussian) limit,
behaves in a well defined way \cite{Cowan:2010js}. The Fisher Information
matrix is connected, through the Cramer-Rao bound
\cite{cramer2016mathematical}, to the full covariance of the model parameters
without the need for computationally expensive Monte Carlo (MC) simulations.
This becomes even more important in the small-sample (Poisson) limit when
asymptotic distributions of the MLR break down. In the past the Poisson limit is the precise regime
in which the Fisher method also breaks down.

Ref.~\cite{Edwards:2017mnf} presents the \textit{equivalent counts} (EC) method
which extends the Fisher formalism to the Poisson regime by condensing the
complexity of calculating exclusion limits and the discovery reach into just two
numbers, equivalent signal and background counts. The success of this method
comes from the mapping between profiled log-likelihood ratios and the
log-Poisson ratio explained in Sec.~\ref{sec:implementation}.  Most importantly the
EC method is designed to be accurate in both the Poisson and Gaussian regimes,
making it a very powerful forecasting tool for a large variety of counting
experiments. In addition to its high accuracy it remains fast and simple to
calculate even when encountering many parameters in the Poisson regime, which
has traditionally been extremely time consuming due to the large number of MCs
required to accurately map the MLR.

\swordfish is a complete and rigorous python package to efficiently manage
and perform all calculations proposed in Ref.~\cite{Edwards:2017mnf} in
addition to some new techniques we introduce here. We provide here a
comprehensive review of the capabilities of \swordfish to calculate the main
quantities of interest, namely exclusion limits and the discovery reach. We
also emphasize a concept we named the \textit{information flux} which acts as a
generalization of signal-to-noise ratio (SNR) by taking into account arbitrary
systematics between background components. 

Traditionally the Fisher information matrix is visualized by viewing its
eigenvalues and eigenvectors as the magnitude and directions of the axes of an
ellipse. Unfortunately this does not capture the full shape of the parameter
space. A major benefit of the Fisher Information matrix is that is can be
treated as a metric on the space of model parameters, allowing for unique
visualization schemes which have been developed in the field of
\textit{information geometry}. We present two main examples: The first is
designed to match as closely as possible the traditional confidence contours
used for parameter reconstruction. We use the geodesic equation as defined by
the Fisher metric to trace the local curvature of the parameter space which is
then easily relatable to a distance scale in terms of standard deviations and
allows for the construction of a non-ellipsoidal contour. The second is a more
generalized visualization scheme using streamlines which represent approximate
$1 \,\sigma$ boxes in the parameter space. When lines become too far apart an
additional line is added to maintain the $1 \,\sigma$ separation. Both of these
schemes are fully contained within \swordfish and are therefore easy and fast
to produce.

A typical problem for both experimental and theoretical physicists is
calculating whether models are discriminable by future experiments. In high dimensional
parameter this is best done by comparison pair-wise comparison of more than 100 million parameter points.
To that end we introduce the \textit{Euclideanized signal} method which approximately maps the
a signal to a new vector which can then be used to calculate the Euclidean distance 
between points. This mapping allows for extremely efficient comparison of a large number of points 
using modern clustering algorithms but only work in Euclidean space.

Indirect and direct dark matter (DM) searches are set to make significant gains
over the next few years with the construction of multiple new experiments. We
here confirm the results EC method by producing approximate DM sensitivity
forecasts for a couple of well known future/current generation experiments, namely
the Cherekov Telescope array (CTA) and Xenon 1T.

\medskip

This paper is structured as follows. In Sec.~\ref{sec:implementation} we
present the interface to \swordfish and provide descriptions of all the
concepts required to understand the EC method\footnote{For more mathematical
details see Ref.~\cite{Edwards:2017mnf}}. Sec.~\ref{sec:Physics} implements
\swordfish projected exclusion limits for both indirect and direct DM searches
before discussing the novel visualizations of the parameter space. 
In Sec.~\ref{sec:Conclusions} we conclude. Additional mathematical details and
verification of the methods can be found in the Appendix.

\section{Structure of package}
\label{sec:implementation}

We here give a basic overview of the package with accompanying descriptions of
the various statistical techniques and approximations used. For more
mathematical details see \ref{apx:Defs} as well as Ref.~\cite{Edwards:2017mnf}.

\subsection{Model definition}

The log-likelihood function of the model that is implemented in \swordfish\ is
a general Poisson likelihood,
\begin{widetext}
\begin{equation}
\label{eqn:model}
  \ln\mathcal{L}_\text{p}(\mathcal{D}| \vect S)
  = \max_{\delta \vect B} \left(\underbrace{\sum_{i=1}^{\nbins} \left( d_i \cdot\ln \mu_i(\vect S,\vect{\delta B}) - \mu_i(\vect S,\vect{\delta B}) \right)}_{(1)}
  - \underbrace{\frac{1}{2}\sum_{i,j=1}^{\nbins} \delta B_i \left(K^{-1}\right)_{ij} \delta B_j}_{(4)} \right)
  \;.
\end{equation}
\end{widetext}
Here, the expectation values $\mu_i$ for the number of events in bins
$i=1,\dots,\nbins$ are given by a sum of the signal $S_i$ and background $B_i$,
as well as background perturbations $\delta B_i$, multiplied by the exposure
$E_i$:
\begin{equation}
  \label{eqn:mu}
  \mu_i(\vect S,\vect{\delta B}) = \Big( \underbrace{
  S_i+B_i}_{(2)}+\underbrace{\delta
  B_i}_{(4)}\Big) \cdot \underbrace{E_i}_{(3)} \;.
\end{equation}
The matrix $K$ describes the covariance of the background perturbations $\delta
B_i$.  Background perturbations are limited to Gaussian variations but
otherwise completely flexible.  Since the background perturbations are usually
not of interest, they are always implicitely profiled out as indicated in the
definition of $\mathcal{L}_\text{p}$, Eq.~\eqref{eqn:model}.

\medskip

The individual terms of the profile likelihood and the expectation values have
the following meaning.
\begin{enumerate}[(1)]
  \item Conventional Poisson log-likelihood (up to a constant term that does
    only depend on data and not affect Frequentist inference).
  \item Linear model, defined through one fixed (`background` or `base`)
    component $\vect B$, the background perturbations $\vect{\delta B}$, and a
    signal component $\vect S^{(k)}$.  The latter can be often a function of
    model parameters $\vect \theta$ (a parameteric background model is instead
    split up in a base part $\vect B$ and its variations $K$).
  \item Exposure.  It is defined such that $\mu_i$ is dimensionless and
    corresponds to expected counts in bin $i$.  The physical units of the
    exposure and the signal and background components is up to the user and
    situation dependent.  Factoring out exposure in the definitions is necessary for the definition of \textit{information flux} further below.
  \item Correlated background perturbations with a covariance matrix $K_{ij}$
    of size $\nbins\times\nbins$.  If the number of bins is large, this
    introduces an equally large number of additional nuisance parameters in the
    problem.  Our approach handels these efficiently.
\end{enumerate}

A few final remarks about the model in Eq.~\eqref{eqn:model} are in order.  The
expectation values $\mu_i$ as defined in Eq.~\eqref{eqn:mu} are not
automatically positive.  For large enough variances $K$, the $\delta B_i$ can
become negative enough to change the sign of $\mu_i$.  The physical constraint
$B_i+\delta B_i\geq0$ cannot be directly implemented in our treatment.  A
straightforward solution to this problem would be to replace $B_i + \delta B_i
\mapsto B_i \exp(\delta B_i / B_i)$.  If we were to do a similar transformation
for the signal model, the resulting model for $\mu_i$ would be positive for any
model parameters.  Such a scenario, a mixture between a Poisson process and a
Gaussian random field, is known as Cox process~\cite{10.2307/2983950}.  However, although
this model has more appealing mathematical properties, it somehow dilutes the
connection with the typical (usually additive) models one encounters in
particle and astro-particle physics.  Furthermore, up to the second derivative
of the model parameters, which is what we are interested in here, both models
would give identical analytic results.

\subsection{Overview}

\begin{figure*}[t]
 \centering
 \includegraphics[width=0.99\linewidth]{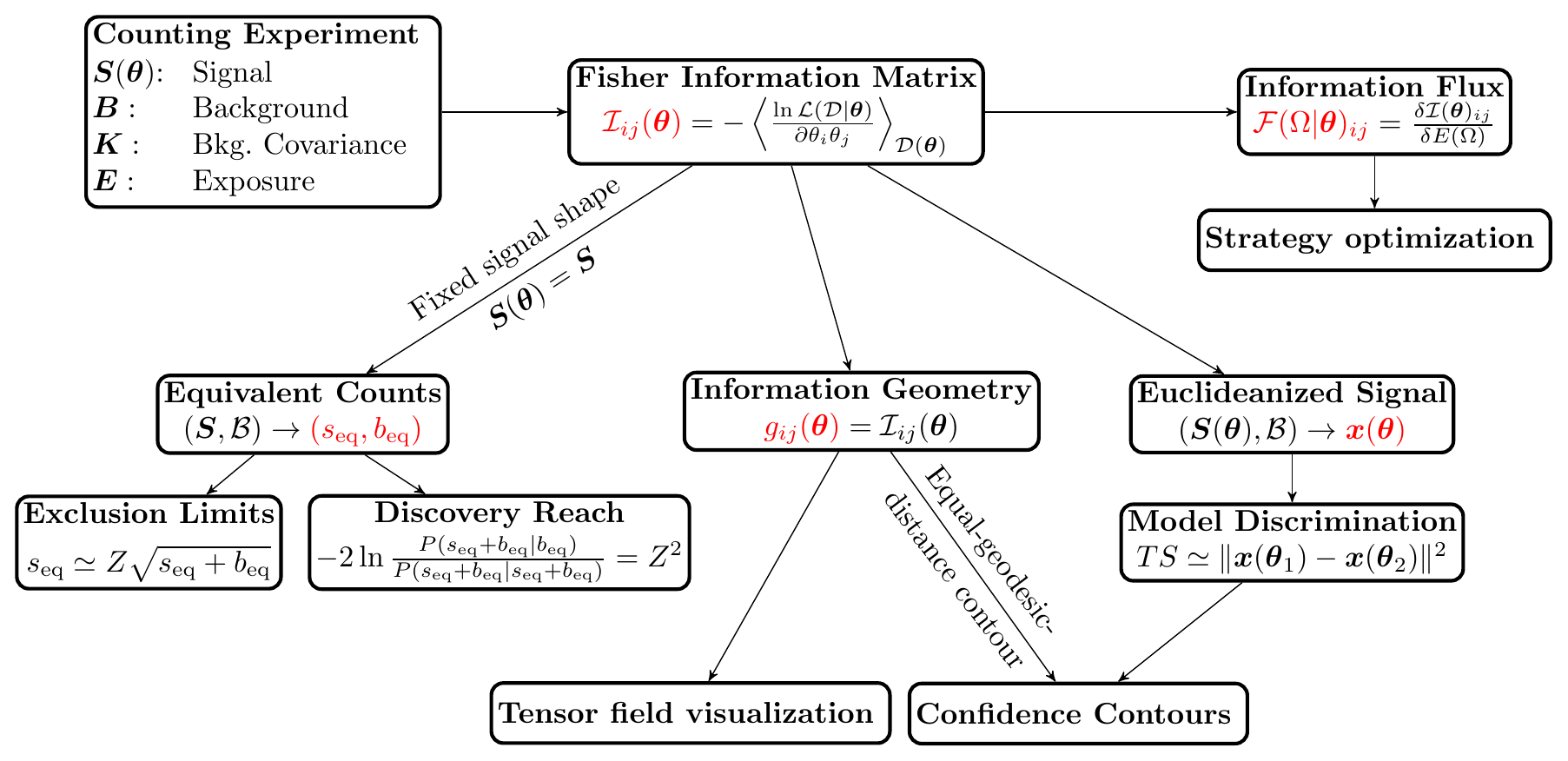}
 \caption{Overview of the various capabilities of the \swordfish\ package. The arrows throughout the diagram indicate the work flow of \swordfish. Starting with the input in the top left corner, a counting experiment within \swordfish is completely described by the signal(s), background(s), background covariance, and exposure as defined in Eq.~\ref{eqn:model}. The primary object is the Fisher Information matrix which then allows for the computation of multiple quantities. For exclusion limits and discovery reach we use the Equivalent Counts method described in \cite{Edwards:2017mnf}. Treating the Fisher matrix as a metric on the parameter space then allows for two visualization schemes, adaptive streamlines and traditional confidence contours. We also introduce here a new technique known as the Euclideanized signal for model discrimination in future experiments and as an alternative route to confidence contours in multi-model parameter spaces. Finally, we allow for fast computation of the Information flux, a generalization of the signal to noise ratio, for strategy optimization.}
 \label{fig:diagram}
\end{figure*}

In Fig.~\ref{fig:diagram}, we provide a conceptual summary of the calculations
that can be performed with \swordfish\ and their relations.  Our `counting
experiment' is fully defined by the (potentially parametric) signal $\vect
S(\vect \theta)$, the base background $\vect B$ and its perturbations $\delta
B$ as described by the covariance matrix $K$, and the exposure $\vect E$.
Based on this model, the Fisher information matrix can be calculated.  The
Fisher matrix can then be used in various ways.  First, for fixed signal shapes
with a free normalization, it is possible to use the equivalent counts method
to forecast exclusion limits and the discovery reach of the implemented
counting experiment.  Second, one generates a field of Fisher information
matrices, which gives rise to a metric in the space of model parameters.  This
metric can be used to derive expected confidence contours for parameter
reconstruction, or it can be visualized directly using adaptive density
streamlines.  Thirdly, the Euclideanized signal method can be used to calculate
the discrimination power of the couting experiment w.r.t.~various signal
benchmark models.  This anasatz can again be used to derive confidence
contours.  Lastly, it is possible to calculate the information flux, which is a
generalization of the signal-to-noise ratio to scenarios where background
systematics can no longer be neglected.

In the rest of this section we will present the various components of
\swordfish\ and show how they are used in practice.  Most of the theoretical
background can be found in Appendix.~\ref{apx:Defs}, as well as in our
technical paper Ref.~\cite{Edwards:2017mnf}.

\subsection{Examples I: Variance, Information Flux, Limits and Discovery Reach}

\begin{figure}[t]
  \centering
  \includegraphics[width=\linewidth]{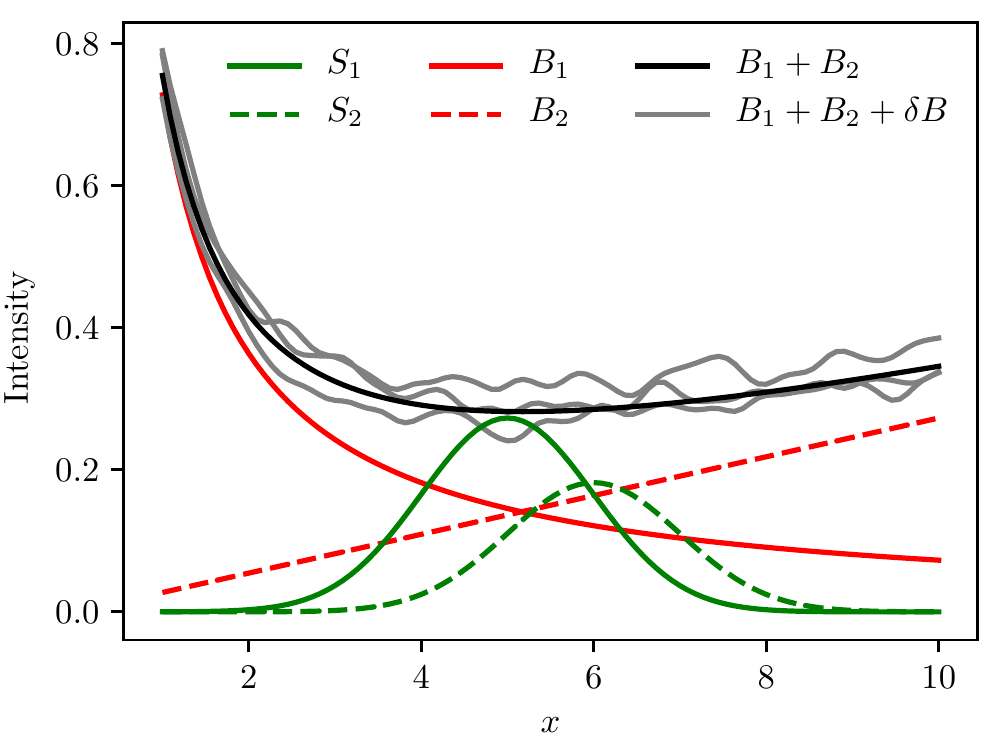}
  \caption{Possible signal and background components adopted in Example I.  We
  also indicate some realizations of the background uncertainties $\delta B$,
  as they are encoded in the covariance matrix $K$, by sampling from the
  corresponding mutlivariate normal distribution.}
  \label{fig:sigs}
\end{figure}

For the purpose of demonstrating the various capabilities of \swordfish, we set
up a simple model with two background components, two different signal shapes,
a covariance matrix for background uncertainties, and flat exposure.  Most of
the code is for the construction of example signal and background spectra.
\begin{lstlisting}
import numpy as np
import swordfish as sf

# Basic grid
x = np.linspace(0, 10, 100)
dx = x[1] - x[0]

# Signal components
S1 = 3.0*np.exp(-(x-5)**2/2.)*dx
S2 = 2.0*np.exp(-(x-6)**2/2.)*dx

# Background components
B1 = 8./x*dx
B2 = 0.3*x*dx

# Background covariance matrix
X, Y = np.meshgrid(x,x)
K = (np.exp(-(X-Y)**2/20.)*0.02**2
    +np.exp(-(X-Y)**2*10.)*0.01**2)

# Exposure
E = np.ones_like(B1)*100.
\end{lstlisting}
The instantiation of \swordfish\ is just a single line, depending only on the
various background components and the exposure.
\begin{lstlisting}
# Instantiate Swordfish
SF = sf.Swordfish([B1, B2], T=[0.1, 0.],
                E=E, K=K)
\end{lstlisting}
We assumed here that there are two background components (first argument),
where the normalization of the first component, $B_1$, is uncertain by $10\%$
(second argument).  Furthermore, the third and fourth argument describe
respectively the exposure and the covariance matrix.  Note that this is just a
convenient wrapper and that internally the background components and their
uncertainties are recast in terms of the total background $B$ and additional
contributions to the covariance matrix $K$.  

\begin{figure}[t]
  \centering
  \includegraphics[width=\linewidth]{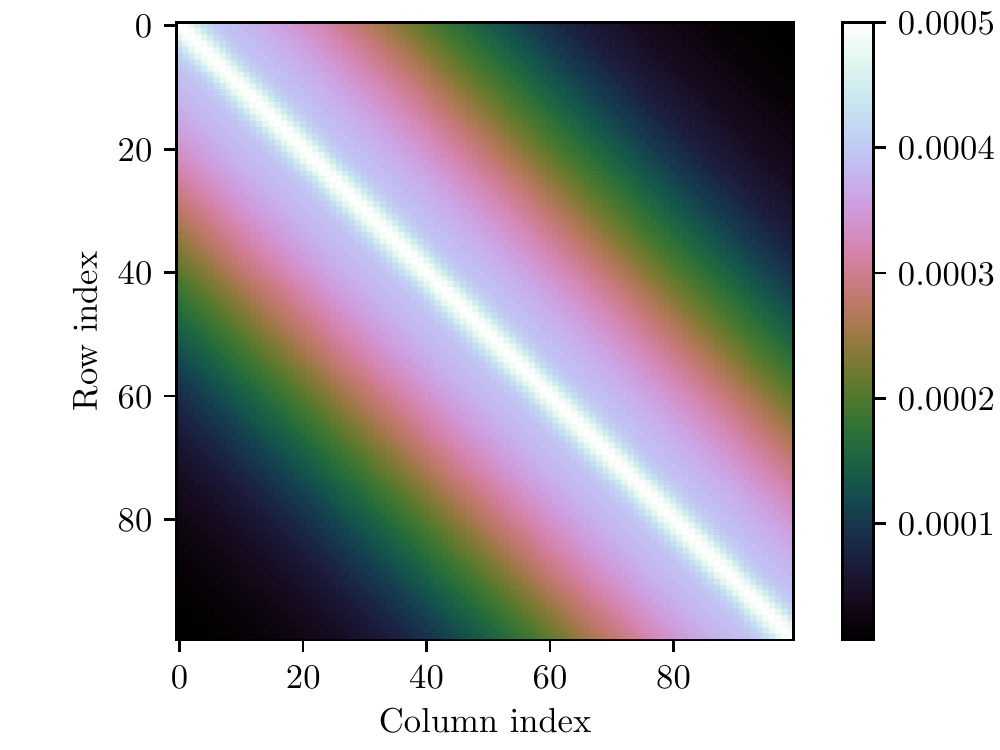}
  \caption{Illustration of the covariance matrix that we adopted in Example I.
  It has a narrow component, as well as a weak broad component.}
  \label{fig:cov}
\end{figure}

In Fig.~\ref{fig:sigs} we show for convenience the two signal and background
components that we defined above.  We also indicate possible realizations of
the background variations that are encoded in the covariance matrix $K$.
Furthermore, we illustrate in Fig.~\ref{fig:cov} the covariance matrix.

\medskip

\paragraph{Covariance and Fisher matrix.} The covariance matrix, $\Sigma$, of
multiple signal components can be calculated by calling the method
\lstinline{covariance}, with a list of signal shapes as argument.
\begin{lstlisting}
# Calculate covariance between multiple 
# signal components
print SF.covariance([S1, S2], S0=2*S1)
>> [[ 0.52 -0.60]
    [-0.60  1.16]]
\end{lstlisting}
More specifically, this corresponds to the covariance matrix for a linear model
with $\vect S(\vect \theta) = \theta_1 \vect S_1+ \theta_2 \vect S_2$, around
values of $\theta_1 = 2$ and $\theta_2 = 0$ (i.e., we assume that there is an
additional contribution of $2\vect S_1$ to the background noise level, which is
provided via the keyword argument \lstinline{S0}).  General non-linear models
can be handled by numerical differentiation as discussed below in
Sec.~\ref{sec:exII}.

Technically, the covariance matrix is derived as matrix inverse of the Fisher
information matrix, $\mathcal{I}$, which can be obtained using the
\lstinline{fishermatrix} method.
\begin{lstlisting}
# Obtain Fisher information matrix for
# multiple signal componets
print SF.fishermatrix([S1, S2])
> [[ 4.74  2.43]
   [ 2.43  2.11]]
\end{lstlisting}
Here, we assumed an expansion around $\theta_1=\theta_2=0$, since the default
value of \lstinline{S0} is zero.

\medskip

\paragraph{Information flux.} With \swordfish, it is possible to calculate the
\textit{information flux}, which we proposed as a generalization of the common
signal-to-noise ratio (SNR) in Ref.~\cite{Edwards:2017mnf}.  It quantifies how
infinitesimal changes of the exposure in each bin affect the variance.  We
hence define the information flux associated with bin $i$ as the partial
derivative of the inverse variance w.r.t.~the exposure in the same bin, $E_i$,
namely
\begin{equation} \mathcal{F}_i \equiv \frac{\partial(1/\sigma^2)}{\partial
E_i}\;.  \end{equation}
Here, we assume a single-component linear signal model, $\vect S(\theta) =
\theta \vect S_1$, and $\sigma^2 \equiv \Sigma_{\theta\theta}$.  Note that in absence
of background uncertainties, this definition simply yields $\mathcal{F}_i =
S_i^2/B_i$, which is indeed the conventional SNR.

\begin{figure}[t]
  \centering
  \includegraphics[width=1\linewidth]{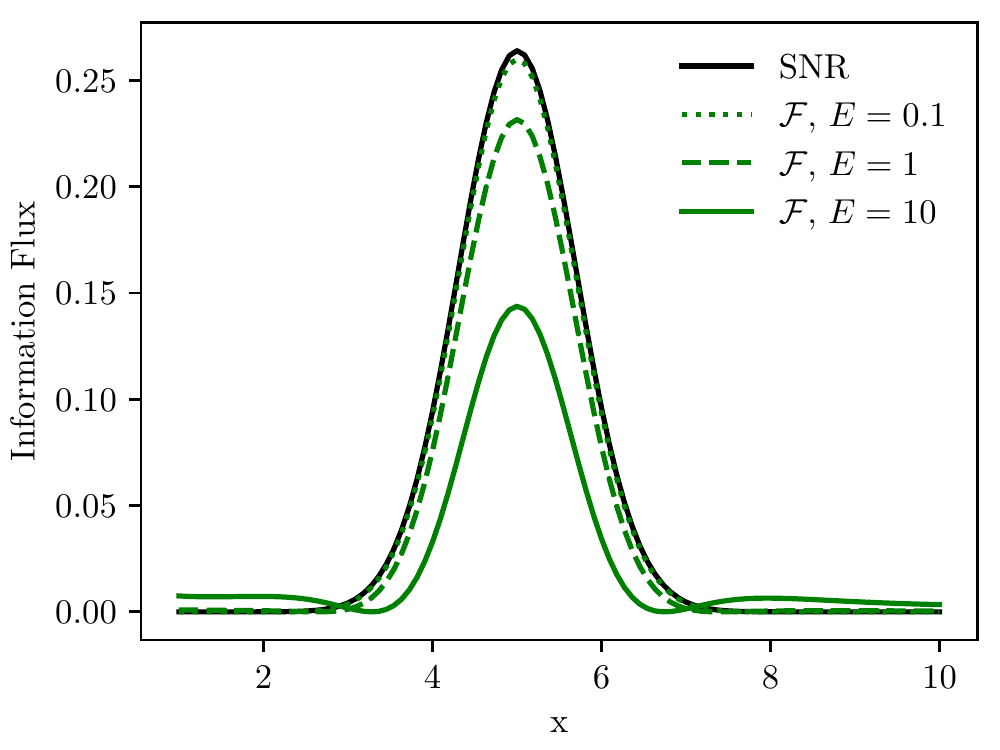}
  \caption{The signal-to-noise ratio of signal $S_1$, compared to the information flux at different values of the exposure $E$.  Here, $E=1$ corresponds to our baseline model.  It is clearly evident that the information flux drops as the measurement enters the systematics dominated regime (for large exposures), and side-bands become more relevant since they facilitate signal background discrimination.}
  \label{fig:F}
\end{figure}

In \swordfish, the information flux for a linear one-component signal model can be
calculated using the \lstinline{infoflux} method, providing the signal shape as
argument (here, the signal is never added to the background noise).
\begin{lstlisting}
# Obtain information flux
print SF.infoflux(S1)
> [9.7e-04, 9.2e-04, ..., 7.2e-5]
\end{lstlisting}
The information flux for $S_1$ is shown in Fig.~\ref{fig:F}, at different
exposure times.  It is evident that for short exposure times, where
observatoins are still limited by statistical noise, the information flux
equals the SNR. For larger exposure times, however, where the measurement
enters the systematics limited regime, the information flux from the main
signal region decreases drastically, while the information flux from
`side-band' regions becomes enhanced.  This is evident in Fig.~\ref{fig:F} for
exposures above $E\sim10$.

\begin{figure}[t]
  \centering
  \includegraphics[width=1\linewidth]{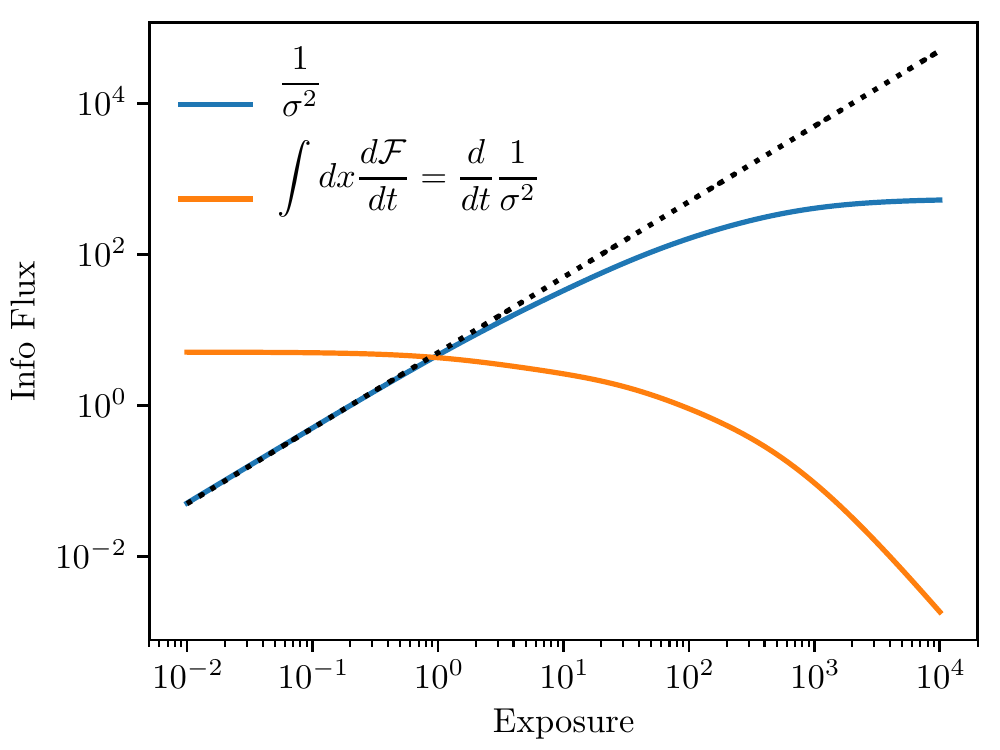}
  \caption{The inverse variance of the $S_1$ normalization as function of the
  exposure.  We also show the integrated information flux, which equals the
  derivative of the inverse variance w.r.t.~the exposure.}
  \label{fig:Fint}
\end{figure}

Summing the information flux over all bins simply yields the exposure
derivative of the inverse variance, $d\sigma^{-2}/dE$, if we pretend that the
exposure is increased simultaneously in all bins (otherwise the integral should
be reweighted accordingly).  Integrating this expression from zero to some
total exposure gives back the inverse variance as function of that total
exposure.  This is illustrated in Fig.~\ref{fig:Fint}.  We indicate also the
case where the inverse variance increases linearly with the exposure, as is the
case in the statistics-limited regime.  Deviations occur for exposures that
enter the systematics limited regime.

\medskip

\paragraph{Equivalent counts method.} In the Gaussian regime, the (co-)variance
of a signal is enough to estimate discovery thresholds and expected upper
limits.  However, this is not the case closer to the Poisson regime.  To obtain
accurate results in this case, we adopt the \textit{equivalent counts method}
that we developed in Ref.~\cite{Edwards:2017mnf}.  It is based on the idea that
the signal $\vect S$, together with the background and its uncertainties
defined in some \lstinline{Swordfish} instance $\mathcal{R}$, can be mapped on
two non-negative numbers,
\begin{align}
  (\vect S, \mathcal{R}) &\to (s_\text{eq}, b_\text{eq})
\end{align}
such that the full profile log-likelihood ratio, is approximated by the
log-Poisson likelihood ratio,
\begin{align}
  -2\ln \frac{\mathcal{L}_\text{p}(\mathcal{D}_\mathcal{A}(\vect S_0)|\vect S_0)}{\mathcal{L}_\text{p}(\mathcal{D}_\mathcal{A}(\vect S_0)|\vect S)}
  \simeq
  -2\ln \frac{P(b_\text{eq}|b_\text{eq})}{P(b_\text{eq}|s_\text{eq}+b_\text{eq})}
  \;.
  \label{eqn:PoissonApprox}
\end{align}
Here, $\mathcal{D}_\mathcal{A}(\vect S)$ refers to `Asimov data'~\cite{Cowan:2010js}, where we set $d_i = \mu_i(\vect S, \vect{\delta B} = 0)$,
and $\vect S_0 \equiv 0$ corresponds to a vanishing signal.  This leads to
median expected limits and discovery thresholds.  We will show the numerical
validity of this approximation in the appendix \ref{apx:validation}.  Based on
the Poisson likelihood ratio, it is then possible to derive accurate estimates
for the discovery reach and expected exclusion limits. 

For a single-component linear signal, $\vect S(\theta) = \theta \vect S$, we
define the \textit{equivalent number of expected signal and background counts}
as,
\begin{equation}
  \label{eqn:si}
  s_{\text{eq}}(\theta) \equiv \frac{\theta^2}{\sigma^2(\theta)- \sigma^2(\theta_0)}
  \;,
\end{equation}
and
\begin{equation}
  \label{eqn:bi}
  b_{\text{eq}}(\theta) \equiv \frac{\theta^2\sigma^2(\theta_0)}{[\sigma^2(\theta)- \sigma^2(\theta_0)]^2}
  \;,
\end{equation}
where $\sigma^2 \equiv \Sigma_{\theta\theta}$ like above.  When calculating
$\sigma^2(\theta)$, the signal is assumed to contribute to the background
noise, whereas we used $\theta_0\equiv0$ to indicate that the corresponding
variance is calculated while setting the signal contribution to the background
to zero.

Although probably not obvious on first sight, these expressions lead to the
exact number of signal and background events for the single-bin Poisson
process.  We use them here to generalize the notion of signal and background
counts to any complex experiment in the form of Eq.~\eqref{eqn:model}. For more
detail on the motivation behind these definitions see \ref{apx:ECmethod}.
Validations of the method can be found in Ref.~\cite{Edwards:2017mnf}.

\medskip

\begin{figure}[t]
  \centering
  \includegraphics[width=1\linewidth]{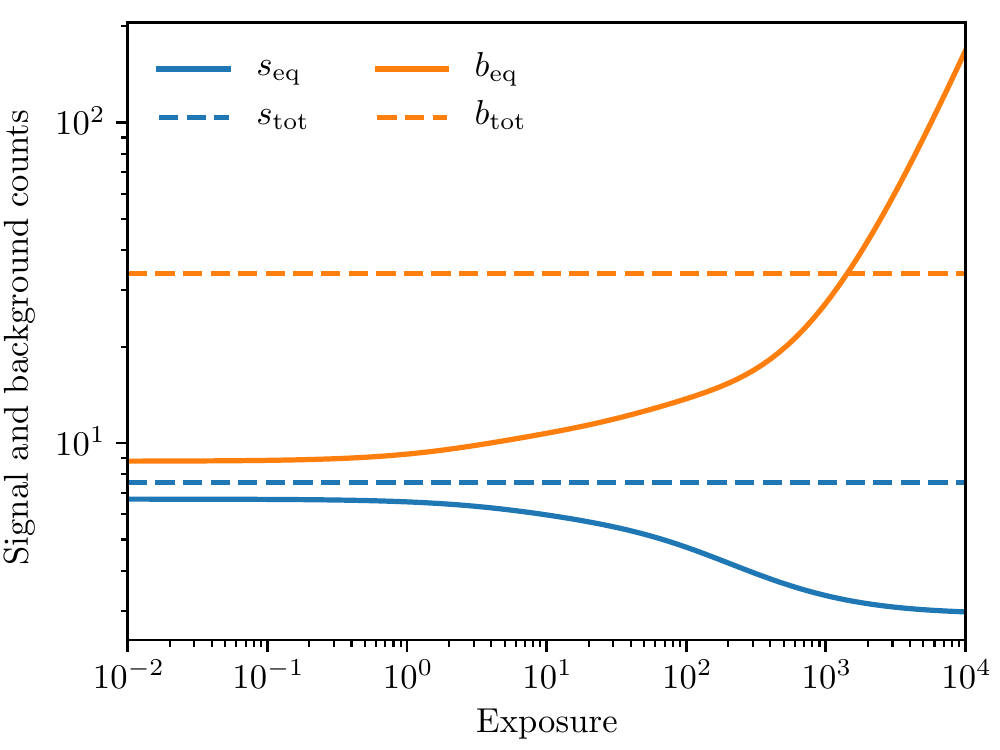}
  \caption{Equivalent signal and background counts for signal $S_1$, as
  function of the exposure.  We clearly see that in the signal and statistics
  limited regime, the number of equivalent signal and background events is in
  general less than the total number of signal and background events.  This is
  expected since only events in the most sensitive signal region contribute.
  However, in the systematics limited regime, the number of equivalent signal
  events drops further, whereas the increasing number of background events
  corresponds to the larger systematic error bars.}
  \label{fig:sb}
\end{figure}

The equivalent signal and background counts are obtained using the
\lstinline{equivalentcounts} method, with the signal shape as argument.  
\begin{lstlisting}
# Obtain equivalent signal
# and background counts
s, b = SF.equivalentcounts(S1)
print s, b
> 6.58 9.15
\end{lstlisting}
The connection between profile log-likelihood and Poisson likelihood will be
discussed further at the end of this subsection.  On the other hand, the
\textit{total} signal and background counts, summed over all bins, can be
calculated with the \lstinline{totalcounts} method. 
\begin{lstlisting}
# Obtain total signal and background
# counts
s, b = SF.totalcounts(S1)
print s, b
> 7.52 33.8
\end{lstlisting}

Before moving on to discuss how the equivalent signal and background can be
used for calculating expected upper limits and the discovery reach, we
illustrate their behaviour in Fig.~\ref{fig:sb}.  Note that the count numbers
are divided by the exposure, to emphasize the non-linear parts of the relation.
For comparison, we also show the total signal and background counts in the
example.  The equivalent signal counts are always somewhat lower than the total
number of signal event.  This is related to the fact that the events in the
tails of the signal peak do not effectively contribute to the overall SNR.  The
number of equivalent signal events furthermore starts to decrease in the
sytematics dominated regime.  On the other hand, the number of equivalent
background events is \textit{much} lower than the total number of background
events.  This comes from the fact that only events around the region where the
signal is significant (the `signal region') actually contribute.  However, in
the systematics-limited regime, the number of equivalent background counts
grows significantly.  This accounts for the fact that the systematic error does
not descrease with additional data.  We note that the concept of equivalent
signal and background events is a useful because (a) it leads to quantiatively
(to within good approximation, see below) correct results for discovery reach
and upper limits, and (b) because it provides a notion for estimating how
close to the Poisson regime a specific observation actually is.

\medskip

\paragraph{Discovery reach and expected limits.} Thanks to the approximate
equivalency in Eq.~\eqref{eqn:PoissonApprox}, it is enough to consider a
one-bin Poisson process.  We just have to insert the equivalent counts as
function of signal normalization $\theta$ to obtain the desired general
results.  The \textit{discovery reach} corresponds to the expected number of
equivalent signal events, $s_\text{\rm eq}$, for which the background-only
hypothesis, $b_\text{\rm eq}$, can be rejected with a median statistical
significance of $\alpha$.  It is can be calculated by solving the implicit
equation\footnote{This approach, based on likelihood ratios and $Z(\alpha)$,
leads to results that are in rather good agreement with the technically
correct treatment which involves partial sums over the Poisson distribution and
makes direct connection with the statistical significance $\alpha$, not
$Z(\alpha)$.  See Ref.~\cite{Edwards:2017mnf} for details.}
\begin{equation}
\label{eqn:disreach}
  -2\ln \frac{P(s_\text{eq}+b_\text{eq}|b_\text{eq})}{P(s_\text{eq}+b_\text{eq}|s_\text{eq}+b_\text{eq})} = Z^2\;,
\end{equation}
where $Z$ is derived from the inverse of the standard normal cumulative
distribution distribution, denoted $F_{\mathcal{N}}$, as
\begin{equation}
  Z(\alpha) \equiv F_{\mathcal{N}}^{-1}(1-\alpha)\;.
  \label{eqn:Z}
\end{equation}

With \swordfish the discovery reach can be calculated with the
\lstinline{discoveryreach} method which takes, besides the signal shape,
the statistical power of the signal discovery, $\alpha$.  
\begin{lstlisting}
# Derive discovery reach
alpha = 2.87e-07  # 5 sigma, one-sided
print SF.discoveryreach(S1, alpha)
> 2.85
\end{lstlisting}
Conversely, \swordfish\ also calculates the expected statistical significance
of a given signal.  This is done by the \lstinline{significance} method, which behaves as inverse of the \lstinline{discoveryreach} method.
\begin{lstlisting}
# Derive stat. significance
print SF.significance(S1*2.85)
>> 2.87e-07
\end{lstlisting}
\medskip

Expected upper limits are, in contrast to the discovery reach, conceptually
somewhat ambiguous.  This is related to the fact that any particular procedure
to derive an upper limit from some data implements a compromise between
correct coverage, practicality and resilience to downward
fluctuations.\footnote{We remind that a puritan's upper limit with perfect
coverage behaviour would yield in the case of a strong downward fluctuation
the empty set as confidence interval.  This is an outcome that experimentalists
like to avoid.}  Our procedure to derive an expected $(1-\alpha)$CL upper limit
is simply to solve the implicit equation
\begin{equation}
  s_\text{eq} = Z\sqrt{s_\text{eq} + b_\text{eq}}\;,
  \label{eqn:ULsb}
\end{equation}
where $Z$ is given by Eq.~\eqref{eqn:Z} above.  For upper limits with $95\%$ or
$99.7\%$ CL, the method leads to good results even in the $b_\text{eq}\to0$
limit, due to some numerical coincidence which is discussed in
Ref.~\cite{Edwards:2017mnf}.  More specifically, we found that the above
procedure leads to results that are in good (usually $<10\%$, in
specific extreme cases up to $40\%$) agreement with median upper limits that we
derived with a fully coverage corrected Monte Carlo for a series of benchmark
scenarios.  

In \swordfish the upper limit is calculated by using the
\lstinline{upperlimit} method, providing the significance level as well as the
signal shape as arguments.
\begin{lstlisting}
# Derive expected upper limits
alpha = 0.05  # 95% CL
print SF.upperlimit(S1, alpha)
> 0.99
\end{lstlisting}

\begin{figure}[t]
  \centering
  \includegraphics[width=1\linewidth]{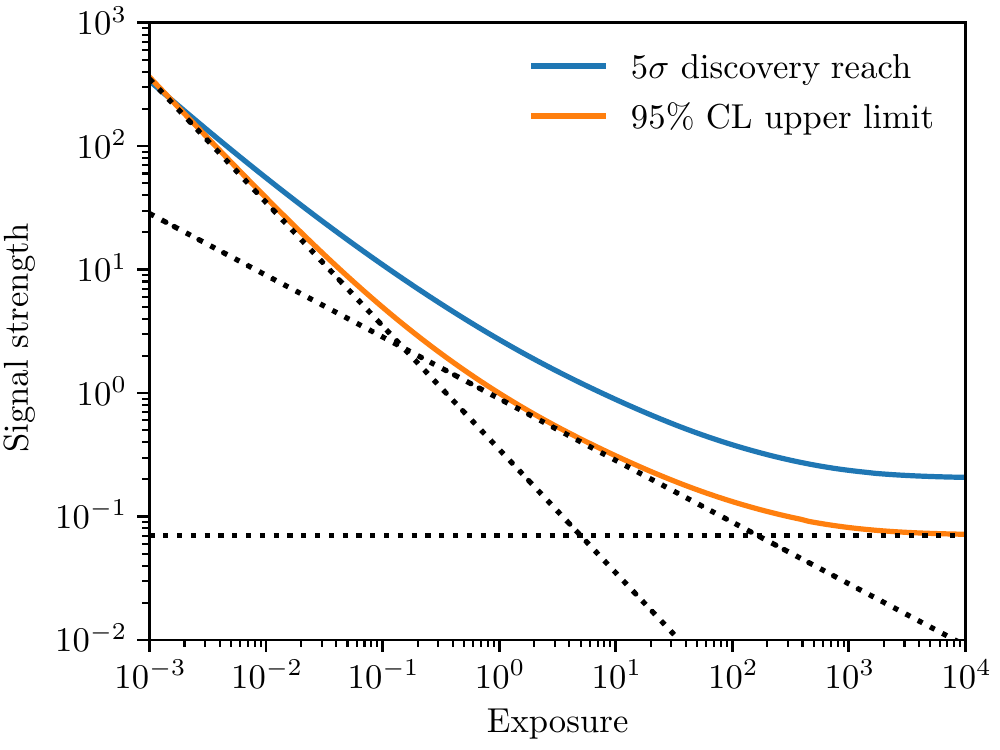}
  \caption{We show projected upper limits on the signal components $S_1$ as
  well as the discovery threshold.  The upper limits show clearly the typical
  behaviour in the signal limited ($\propto E^{-1}$), background limited
  ($\propto E^{-0.5}$) and systematics limited ($\propto\text{const}$) regime.}
  \label{fig:limits}
\end{figure}

In Fig.~\ref{fig:limits}, we show the expected upper limits and discovery reach
for signal $S_1$ as function of the exposure.  We also indicate the expected
scaling of the upper limit in the signal limited ($\propto E^{-1}$), background
limited ($\propto E^{-1/2}$) and systematics limited ($\propto \text{const}$)
regimes.  These are all reproduced properly.

\medskip

\paragraph{Euclideanized signal method.}  The model discrimination power of an
experiment is often quoted as the significance level $z$ (in terms of standard
deviations) at which two benchmark signals $\vect S_1$ and $\vect S_2$ can be
distinguided.  Here, $z$ can be estimated from the expected profile
log-likelihood ratio,
\begin{align}
  \TS = -2\ln \frac{\mathcal{L}_\text{p}(\mathcal{D}_\mathcal{A}(\vect
    S_2)|\vect S_1)}{
    \mathcal{L}_\text{p}(\mathcal{D}_\mathcal{A}(\vect S_2)|\vect S_2)}\;,
  \label{eqn:TScorrect}
\end{align}
where $\mathcal{D}_\mathcal{A}$ denotes again Asimov data as above.  Using the
Fisher approximation, this TS value can be approximated by
\begin{align}
  \TS \simeq \sum_{ij} \Delta S_i D^{-1}_{ij} \Delta S_j\;,
  \label{eqn:xy}
\end{align}
where $\Delta \vect S\equiv \vect S_1 - \vect S_2$ is the signal difference,
and the total (statistic plus systematic) covariance matrix is given by
\begin{equation}
  D_{ij} \equiv K_{ij} + \delta_{ij} \frac{S_{2,i}+B_i}{E_i}\;.
  \label{eqn:D}
\end{equation}

The interpretation of the TS value in units of standard deviations is context
dependent.  We will give here one very typical example.  Suppose $\vect S_1$
and $\vect S_2$ are part of a parametric signal model, $\vect S_1 = \vect
S(\vect\theta_1$) and $\vect S_2 = \vect S(\vect\theta_2)$, and that
$\vect\theta$ has $k$ relevant components.  Furthermore, suppose now that
$\vect \theta_1$ are the parameters of the simple null hypothesis that we want
to test,  and the composite alternative hypothesis that we want to discriminate
against is that $\vect \theta$ can aquire any value.  The above TS value
corresponds then approximately to the median TS value that we would measure in
repeated experiments if the true model parameters were $\vect\theta_2$ (we use
as data the Asimov data set corresponding to $\vect\theta_2$, in which case the
maximum-likelihood estimator of $\vect\theta$ in the alternative hypothesis
would be simply $\vect\theta_2$).  Since the alternative and null hypothesis
are nested, and differ in their degrees of freedom by $k$, we can assume that
the TS value is approximately $\chi^2_k$ distributed, with $k$ degrees of
freedom~\cite{Wilks:1938a, Cowan:2010js}.  
The corresponding threshold value for a $(1-\alpha)$CL contour is then given by 
\begin{equation}
\label{eqn:Ychi2}
  Y_k^2(\alpha) = F_{\chi^2_k}^{-1}(1-\alpha)\;.
\end{equation}
For instance, in the case of two parameters, $k=2$, the $68.7\%$CL or
$95.3\%$CL contours correspond to $Y_{k=2}^2=2.32$ and
$Y_{k=2}^2=6.12$ respectively.

\medskip

Calculating the pair-wise TS-values for $N$ different signals would require
$N^2$ matrix inversions, which is prohibitive if the number of points is large
(say, $N\sim 10^6$, which is not a large number in the context of global
scans).  The main idea of the \textit{Euclideanized signal method} is to
reduces this to just $N$ matrix inversions.  To this end, we define the
\textit{Euclideanized signal}, which constitutes a mapping of the signal and
background onto vectors of length $\nbins$,
\begin{align}
  (\vect S_1, \mathcal{B}) \to \vect x_1
  \quad\text{and}\quad
  (\vect S_2, \mathcal{B}) \to \vect x_2\;,
\end{align}
such that to good approximation the TS value can be replaced by the Euclidean distance between $\vect x_1$ and $\vect x_2$,
\begin{align}
  \TS \simeq \lVert\vect x_1 - \vect x_2 \rVert^2\;.
  \label{eqn:TSeuc}
\end{align}

The mapping depends on the specific noise level of the background as well as
the specified background uncertainties.  It is defined in the appendix,
Eq.~\eqref{eqn:eucsig}.  Here, we just list a few limiting cases.  In the
background-limited regime where systematics are neglected, we have $x_i \simeq
S_i \cdot \sqrt{E_i/B_i}$.  In the systematics limited regime, the expression
looks like $x_i = \sum_j (K^{-1/2})_{ij} S_j$.  In both cases, the signal
enters just linearly, since it does not contribute to the background noise.
This is however different in the signal limited regime, which requires some
extra care.  In this limit, we obtain $x_i \simeq 2\sqrt{E_i S_i}$, where the
$2$ is a fudge factor that compensates for the fact that only the square-root
of the signal appears.  However, one can show that the latter implies to lowest
order in $\Delta S_i/\sqrt{\bar S_i}$ that $\TS \approx \sum_i \Delta
S_i^2/\bar S_{i}$, with $\bar {\vect S} \equiv \frac12 (\vect S_1 + \vect
S_2)$.  We show in appendix~\ref{apx:validation} for a series of randomized
models that the mapping works accurately enough for typical applications.

In \swordfish, this mapping is implemented in the \lstinline{euclideanize}
method.  The TS-value between the signals $\vect S_1$ and $\vect S_2$
can be then calculated as shown in the following example.
\begin{lstlisting}
# Euclideanized signal
x1 = SF.euclideanize(S1)
x2 = SF.euclideanize(S2)
TS = ((x1 - x2)**2).sum()
# This is the same as:
# TS = SF.TS(S1, S2)
print TS
>> 1.38
\end{lstlisting}

\medskip

\paragraph{Profile log-likelihood.} In order to double-check the accuracy of
the equivalent counts and equivalent signal methods, \swordfish\ provides a
method to calculate the profile likelihood in Eq.~\eqref{eqn:model} directly.
This is done by calling the method \lstinline{lnL}, which returns the value of
the profile log-likelihood,
$\ln\mathcal{L}_\text{p}(\mathcal{D}_\mathcal{A}(\vect S_0)|\vect S_1)$.
Internally, this profile log-likelihood is maximized w.r.t.~the background
perturbations $\vect{\delta B}$ by using the L-BFGS-B algorithm~\cite{morales2011remark}.  In
the following example we calcualte a profile likelihood ratio, and compare it
with the results from the equivalent counts and the Euclideanized signal
approximations.

\begin{lstlisting}
# Exact profile log-likelihood ratio
S0 = np.zeros_like(S1)  # Zero signal
TS_L = -2*(SF.lnL(S0, S1) 
            -SF.lnL(S0, S0))

# Poisson likelihood ratio,
# using equivalent counts
s, b = SF.equivalentcounts(S1)
TS_P = 2*(s-b*np.log((s+b)/b))

print TS_L, TS_P
> 3.28 3.25
\end{lstlisting}

\begin{lstlisting}
# Exact profile log-likelihood ratio
TS_L = -2*(SF.lnL(S1, S2)
            -SF.lnL(S1, S1))

# Euclideanized signal
x1 = SF.euclideanize(S1)
x2 = SF.euclideanize(S2)
TS_E = ((x1-x2)**2).sum()
print TS_L, TS_E
> 1.43 1.38
\end{lstlisting}

We find that for these specific cases the results are in good agreement with
each other.  Comparisons with a large number of randomized models can be found
in appendix~\ref{apx:validation}.

\subsection{Examples II: Information Geometry, Confidence Contours, Distance
sampling, Validation}
\label{sec:exII}

All examples in the previous subsection were related to signals with a fixed
shape and free normalization.  However, often signals have a general parametric
form.  \swordfish\ provides several methods to handle parameteric signals.  The
typical use case it to derive projected confidence contours.

As an example signal for this subsection, we define a periodic signal with two
parameters for the normalization, $a$, and for the phase offset, $b$,
respectively.
\begin{lstlisting}
# Parametric signal
S = lambda a, b: 
      b*np.sin(x+0.5*a+0.1*b)**2*dx
\end{lstlisting}

\medskip

\paragraph{Covariance and Fisher matrix.} Parametric signal shapes can be
handled by the methods \lstinline{fishermatrix} and \lstinline{covariance},
provided they are linearized before by using the method \lstinline{linearize}.
This can happen for instance by numerical differentiation.  Estimating the
covariance of the above model paraters $a$ and $b$ can then be done as shown in
the following example.
\begin{lstlisting}
# Calculation of covariance matrix
a0, b0 = 2, 6
gradS, S0 = SF.linearize(S, [a0, b0])
print SF.covariance(gradS, S0 = S0)
>> [[0.26, -0.50], [-0.50, 2.44]]
\end{lstlisting}
This method cannot be used if higher-order derivatives of the signal play a
significant role.  In that case, some of the methods that we discuss in the
following section can be used instead.

\medskip

\paragraph{Information geometry and confidence contours.}  The Fisher
information matrix defines a \textit{metric} on the space of model parameters
(which is the topic of \textit{information geometry}).  Expected confidence
regions for parameter reconstruction happen to correspond to
equal-geodesic-distance contours w.r.t.~this metric.  \swordfish\ provides
methods to construct confidence contours by internally solving the geodesic
equation.  However, it also provides a more general way of visualizing the
Fisher information metric using adaptive-density streamlines.  In some sense,
the streamline visualization represents all possible confidence contours
simultaneously, and is hence more general.

The Fisher information metric defines a tensor field, which can be generated by
using the \lstinline{getfield} method.  This method, as well as the
visualization methods, are currently only implemented for signal models with two
parameters. It takes as arguments the signal model, as well as the two lists
that specify the grid on which the tensor field is evaluated.
\begin{lstlisting}
# Generation of tensor field
alist = np.linspace(0.1, 10, 40)
blist = np.linspace(0.1, 10, 40)
TF = SF.getfield(S, alist, blist)
\end{lstlisting}

\begin{figure}[t]
  \centering
  \includegraphics[width=0.99\linewidth]{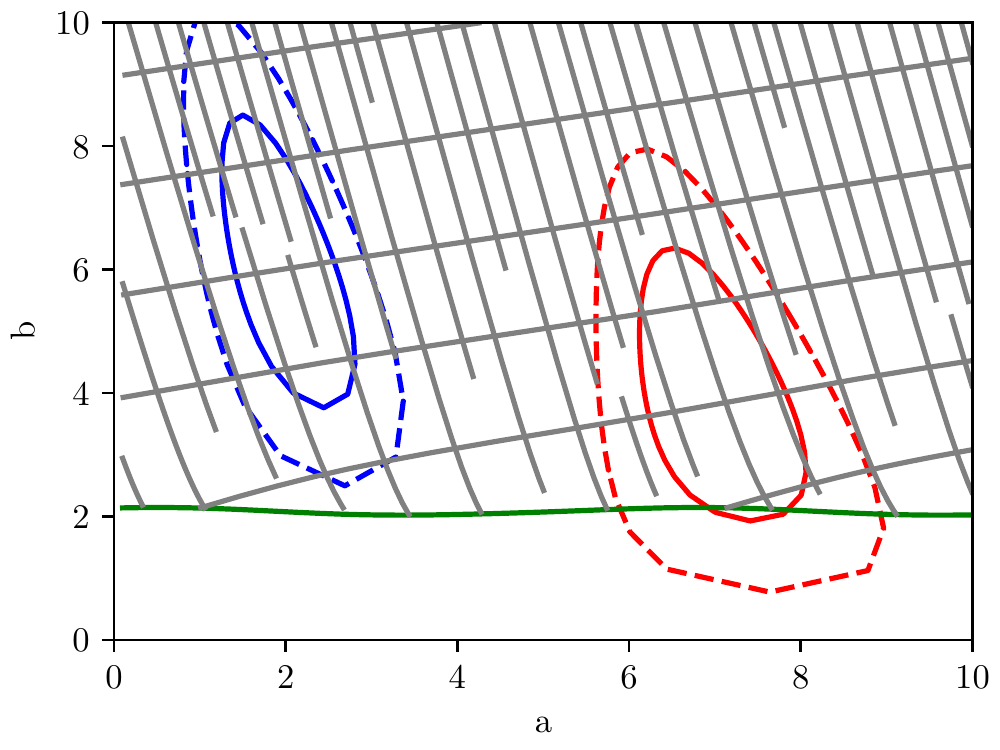}
  \caption{\textit{Gray lines:} Visualization of the Fisher information metric,
    using streamlines.  The line orientation corresponds to the major and minor
    axes of the Fisher information metric (or covariance matrix).  The line
    density changes such that the distance between parallel lines is $\sim
    1\sigma$ (see text for details).  A larger signal normalization $b$ enables
    a better reconstruction of the signal position $a$.  \textit{Red and blue
    lines:} Equal geodesic distance contours, representing expected $1\sigma$
    and $2\sigma$ confidence regions.  Note that the confidence regions
    constructed with this local method do \emph{not} account for the periodic
    structure of the example signal model, which actually leads to disconnected
    confidence regions (see Fig.~\ref{fig:geodesics} for comparison).
    \textit{Green line:} Expected exclusion limits on $b$ (for fixed $a$) at $95\%$CL.}
  \label{fig:flow}
\end{figure}

Based on the tensor field, confidence contours can be generated using the
\lstinline{contour} method, which takes as argument both the central point of
the contours as well as a distance.  
\begin{lstlisting}
# Confidence contour visualization
# (equal geodesic distance contours)
a0, b0 = 6, 2  # center
d = 2.32  # distance
TF.contour([a0, b0], d, color='red')
\end{lstlisting}
This code generates a confidence contour with geodesic distance
$d=2.32$ around the indicated central point $(a_0, b_0)$.  This corresponds to a
$68.3\%$CL region.  Examples for confidence contours generated this way are
shown in Fig.~\ref{fig:flow}

Internally, the method first generates Christoffel symbols by numerical
differentiation of the tensor field.  Second, it shoots geodesics of the
specified length from the central point in a number of directions, and finally
connects the endpoints.  More details can be found in appendix
\ref{subsec:Geodesic}.

\medskip

Instead of showing confidence contours for specific benchmark points, it also
useful to directly visualize the underlying Fisher information metric field.
In fact, this can lead to a more complete, although somewhat unusual,
representation of experimental abilities than specific confidence contours
could do.  A large number of visualization techniques exist for tensor
fields~\cite{laidlaw2009visualization}.  In \swordfish, we implemented some
variation of adaptive-density streamline visualization (e.g.,
Ref.~\cite{tchon2004visualizing}) which works for 2-dim parameter spaces.

Adaptive-density streamline plots are straightforward to interpret:  The
direction of the streamlines corresponds exactly to the major and minor axes of
the Fisher information metric (and hence of the covariance matrix).  The
distance between two parallel streamlines corresponds \emph{approximately} to
$1\sigma$ in the direction perpendicular to the streamlines.  The latter
condition is realized by adding or removing lines as necessary.

In \swordfish, streamlines are generated by first generating
\lstinline{VectorField} objects from the tensor field, using the
\lstinline{VectorFields} method.  Streamlines are then generated with the
method \lstinline{streamlines}.  An example is shown in Fig.~\ref{fig:flow}. 
\begin{lstlisting}
# Streamline visualization
vf1, vf2 = tf.VectorFields()
vf1.streamlines(color='0.5')
vf2.streamlines(color='0.5')
\end{lstlisting}

The streamline generation is an iterative process.  First, two perpendicular
streamlines are drawn starting from a seed location.  Starting from random
points on the existing streamlines, it is then checked whether a parallel line
at around $1\sigma$ distance exists.  If not, it is added and extended until it
gets too close to an already existing streamline.  Further details can be found
in appendix~\ref{apx:infoGeom}.

\medskip

\paragraph{Sampling and confidence contours.}  The above visualization of
confidence contours using equal geodesic distance contours is a local method,
and hence does not correctly treat multi-modal confidence regions.
Furthermore, it would break down when the Fisher information matrix is
singular, which often happens for high-dimensional models that are not
sufficiently constrained by data.  

\begin{figure}[t]
  \centering
  \includegraphics[width=0.99\linewidth]{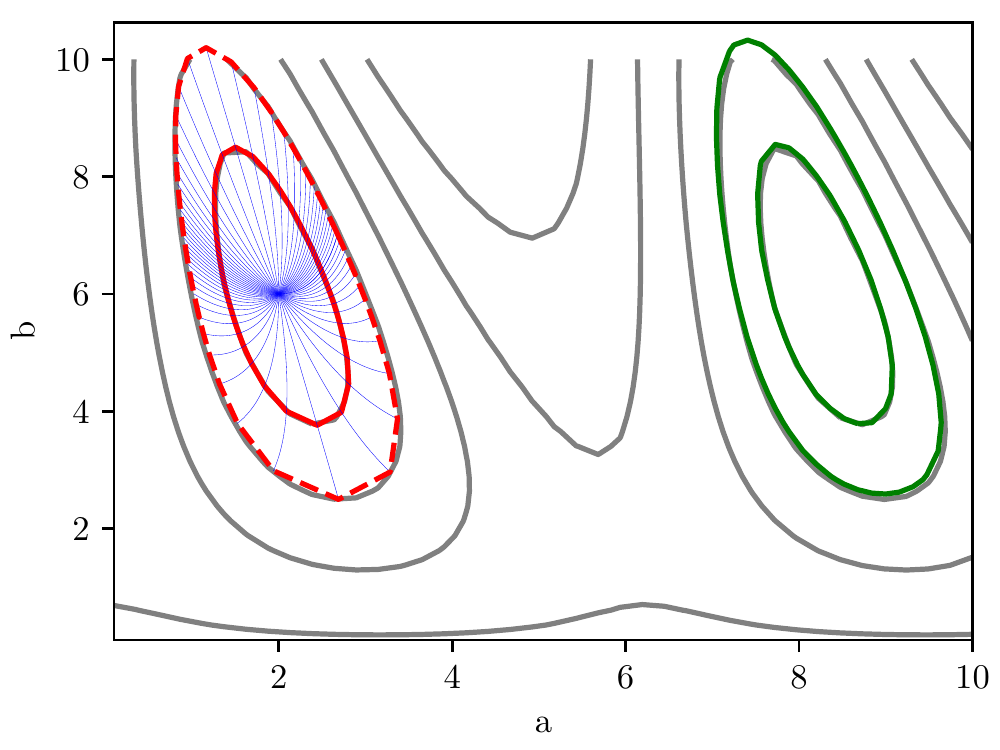}
  \caption{\textit{Gray lines}: Confidence contours constructed by using the
    equivalent shape method, which also works for multi-model confidence
    regions.  \textit{Red lines:} Confidence regions from the equal geodesic
    distance contours (the blue lines show the geodesics used for constructing this
    region).  \textit{Green lines:} Confidence regions construced using the
    \textsf{minos} algorithm of \textsf{minuit}.}
  \label{fig:geodesics}
\end{figure}

In the following example, we first generate a distance field using the
\lstinline{TS} method, which adopts the above Euclideanized signal method to
approximate the profile log-likelihood.
\begin{lstlisting}
# Confidence contour visualization
# (distance measure sampling}
a0, b0 = 3, 5
Sd = S(a0, b0)
TSfield = np.array([[SF.TS(S(a,b),Sd) 
  for a in alist] for b in blist])
plt.contour(alist, blist, TSfield**0.5,
     levels = [1, 2, 3], colors='0.5')
\end{lstlisting}
The resulting contours are shown in Fig.~\ref{fig:geodesics}.  They correctly
account for the multi-modal structure of the confidence contours, and agree
with the geodesic distance contours where they overlap.

\medskip

\paragraph{Validation with numerical likelihood maximization.}  In particle and
astroparticle physics, a very common approach to handling profile log-likelihoods
and related inference problems is to make use of the C++ numerical minimizer
\texttt{minuit}.  In order to validate the approximation schemes used in
\swordfish, it is for any given model possible to directly generate a
\lstinline{Minuit} instance (part of the Python package \lstinline{iminuit}).
This can then be used to derive confidence contours etc and compare with
\swordfish\ results.  This is done using the factory method
\lstinline{getMinuit}, where as arguments the parametric signal as well as the
true signal parameters have to be provided.  Note that the profiling over
background perturbations, $\vect{\delta B}$, is done not by \texttt{minuit} but
by calling L-BFGS-B, as described above for the \lstinline{lnL} method
(L-BFGS-B directly uses analytic gradient information, which leads to a very
significant speed-up).  In the following example, we show how confidence
contours can be generated with \lstinline{Minuit}.
\begin{lstlisting}
# Generate Minuit instance
# of Swordfish model
a0, b0 = 6, 2
M = SF.getMinuit(S, [a0, b0])
M.migrad()
M.minos()
X, Y, TSfield = M.contour(
        "x0", "x1", bound = 3.)
TSfield = np.array(TSfield)
TSfield -= TSfield.min()  # zero
plt.contour(X, Y, TSfield**0.5,
      levels=[1, 2, 3], colors='green')
\end{lstlisting}
The resulting contours are shown in Fig.~\ref{fig:geodesics}, and agree with
contours generated using the geodesic equation or the distance sampling mehtod.

\section{Physics Examples}
\label{sec:Physics}

We will now discuss two typical examples from direct and indirect searches for
dark matter signals.clear  The main purpose of these examples is to illustrate
how the various possible inputs of \swordfish\ connect to physical fluxes and
the exposure, and how various statements about model parameters can be derived
in practice.  Our goal was to keep the examples as simple as possible, and not
to exactly reproduce published results.  However, all observed deviations can
be entierly attributed to the different physics assumptions that are made.  We
showed already above that \swordfish\ reproduced results obtained from the
profile-likelihood method.

\subsection{Galactic center dark matter searches with a CTA-like experiment}

Our first example is inspired by the upcoming Cherenkov Telescope Array (CTA),
and we will largely follow a simplified version of the ON-OFF analysis
discussed in Ref.~\cite{Silverwood:2014yza}.  The idea is to measure the
gamma-ray intensity in a `ON' region close to the Galactic center, and to
compare it with the intensity in an `OFF' region that is somewhat further away.
If no difference in the intensity is observed, upper limits on a dark matter
contribution (which would predominantly contribute to the ON region) can be
obtained.

We assume throughout dark matter annihilation into $\bar bb$ final states, and
an Einasto profile and ON/OFF regions with the same parameters as assumed in
Ref.~\cite{Silverwood:2014yza}.  The J-values and
the angular size of these regions are: $J_\text{ON} = 21\times 10^{21} \rm
GeV^2 cm^{-5}$, $J_\text{OFF} = 21\times 10^{21} \rm GeV^2 cm^{-5}$,
$\Omega_\text{ON} = 3.2\times10^{-3}$, and $\Omega_\text{OFF} =
3.2\times10^{-3}$.  The cosmic-ray electron background is assumed to follow a
broken power law $dN_{e^-}/dE = 1.17\times 10^{-11}
\left(\frac{E}{\mathrm{TeV}}\right)^{-\Gamma} \rm cm^{-2} s^{-1}$, where
$\Gamma = 3.0$ for $E<1$ $\mathrm{TeV}$, and $\Gamma = 3.9$ for $E>1$
$\mathrm{TeV}$.   The cosmic-ray proton flux is assumped to follow a power law
$dN_p/dE = 8.73\times 10^{-9} \left(\frac{E}{\mathrm{TeV}}\right)^{-2.71} \rm
cm^{-2} s^{-1}$, and we assume a $99\%$ rejection efficiency and a factor 3
shift in energy for reconstructed protons (see Ref.~\cite{Silverwood:2014yza}
for details).  The Galactic diffuse emission is neglected for simplicity
(although it likely \textit{does} play a significant role, and should not be
neglected in any real analysis or projection).  We sum both electron and
proton contributions to a single isotropic component, and assume that its
normalization is uncertain by $50\%$, but identical in both the `ON' and
`OFF' regions.   Besides that, we do not adopt any additional systematic
background uncertainties.  Lastly, we adopt the same exposure (now somewhat
outdated) curves as in Ref.\cite{Silverwood:2014yza}.
\medskip

\begin{figure}[t]
  \centering
  \includegraphics[width=0.99\linewidth]{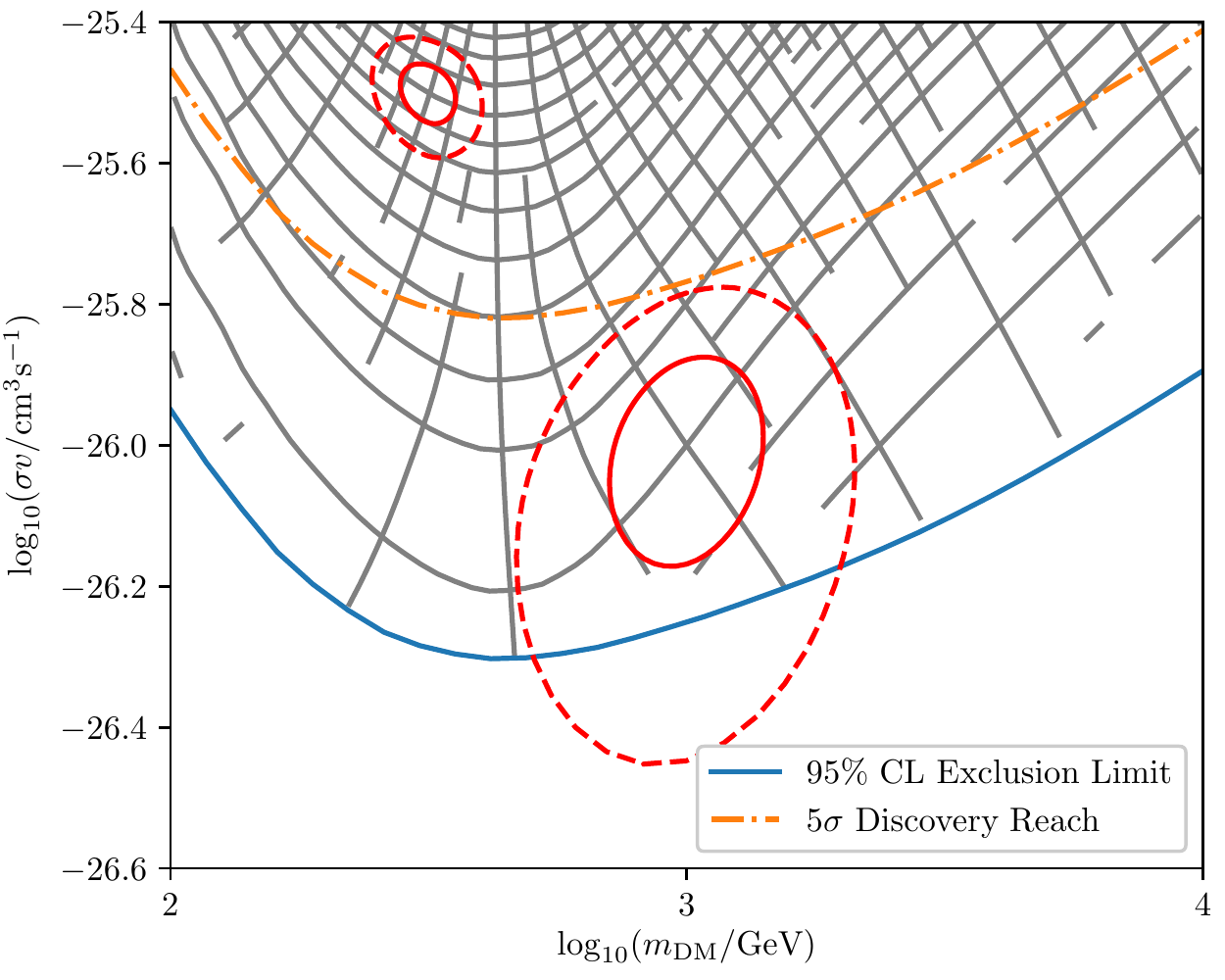}
  \caption{Blue solid line and orange dot-dashed show the $95\%$CL exclusion
  limit and $5\sigma$ discovery reach respectively. We also plot two examples
of $1\sigma$ and $2\sigma$ contours. The grey lines are $1\sigma$ streamline
visualisation, as described in the previous sections.}
  \label{fig:CTA}
\end{figure}

The actual implementation of the present example can be found in the
\texttt{jupyter} notebook \texttt{swordfish\_ID.ipynb}.  In
Fig.~\ref{fig:CTA}, we show the $95\%$CL projected upper limits, the
$5\sigma$ discovery threshold, example reconstruction contours (68.3\% and
95.4\% CL), as well as the streamline visualization of the Fisher information
metric.  As it is clear from the code, all this is relatively straightforward
to obtain once the instrumental details are coded up.

\subsection{Xenon-1T}

Direct detection experiments have, over the past few years, gained several
orders of magnitude in sensitivity over a wide range of DM masses. Sensitivity
at multiple mass scales is achieved mainly through the use of different target
elements allowing various detectors to probe from keV to GeV recoil energies.
The use of many smaller scale direct detection experiments with low backgrounds
have been discussed in the literature recently \cite{Battaglieri:2017aum}
providing excellent motivation for a fast and simple way calculate the
sensitivity of of these future experiments given a physical dark matter model.
Direct detection experiments have, over the past few years, gained several orders of magnitude in sensitivity over a wide range of DM masses. Sensitivity at multiple mass scales is achieved mainly through the use of different target elements allowing various detectors to probe from keV to GeV recoil energies. The use of many smaller scale direct detection experiments with low backgrounds have been discussed in the literature recently \cite{Battaglieri:2017aum} providing excellent motivation for a fast and simple way calculate the sensitivity of of these future experiments given a physical dark matter model. We here implement a simplified calculation of the sensitives for the Xenon-1T experiment.

The Xenon collaboration have recently published their latest results from $35636\mathrm{\, kg\, days}$ of exposure \cite{Aprile:2017iyp}. The only way to reasonably replicate the results of Ref.~\cite{Aprile:2017iyp} is to fully simulate the detector response and signal, tuning your parameters to fit the few number of events in the signal region. Here, we instead calculate the signal distributed over only the primary photon signal (S1). Since Fig.~3 of Ref~.\cite{Aprile:2017iyp} provides the background components as a function of S1 we are able to directly extract these and consider how our DM signal would be distributed in the same plane. Typically a DM signal in a DD experiment is expressed as a recoil spectrum given by 
\begin{equation}
\label{eqn:dRdER}
\frac{dR}{dE_R} = \frac{\rho_0\xi_T}{2\pi m_{\text{DM}}} \frac{g^2 F_T^2(E_{R})}{(2m_TE_{R} + m^2_{\text{med}})^2}\eta\left(v_{\text{min}}(E_R)\right)~,
\end{equation}
where $\rho_0$ is the dark matter density at earth which we take to be $0.3\mathrm{\,GeV\,cm}^{-3}$, $m_{\text{DM}}$ is the dark matter mass, $m_\text{med}$ is the mediator mass, $F_T^2(E_{R})$ is the recoil form factor, and $m_T$ is the mass of the target isotope. We use a common approximation $F_T^2(E_{R}) = \exp(-(r_n\sqrt{2m_TE_R})^2/3)$ for the recoil form factor \cite{lewin1996review}. To match the known backgrounds as a function of S1 we use the change of variables $\frac{dR}{dS1} = \frac{dR}{dE_R}\frac{dE_R}{dS1}$, for which we need a reasonable approximation of $\frac{dE_R}{dS1}$.

For our approximation we take red line in the bottom panel of Fig.~2, Ref~.\cite{Aprile:2017iyp}, to give us the median expected value for the S2 component given a value of S1. We then assume that the points at which the contour lines of constant energy cross the median expected value of S2 approximately describe the typical recoil energy for a given S1. We simply compute the derivative and use the the nuclear recoil efficiency from Fig.~1 along with a factor of $0.475$ to take into account the fact that we are only interested in signals in the reference region. 

To check our approximation we integrate over the entire distribution of S1 ($3-70 \mathrm{\,PE}$) for the benchmark scenario presented in Ref~.\cite{Aprile:2017iyp}, namely a $50 \mathrm{\, GeV}$ DM particle with $\sigma = 10^{-46} \mathrm{\,cm^{-2}}$. Our approximation agrees well with the $0.82$ presented, finding that we get $0.83$ events in the reference region\footnote{To compare with the results of the Xenon1T collaboration we took the mass of the mediator to be very large so it no longer plays a role, $m_{\text{med}} = 10^5 \mathrm{MeV}$}. Note here we take $\sigma = \frac{g^2\mu}{\pi m_{\text{med}}^4}$ where $\mu=\frac{m_{\mathrm{DM}}m_p}{m_{\mathrm{DM}} + m_p}$ and $m_p$ is the mass of the proton. Through the same procedure we sum the backgrounds and compare our total signal in the reference region which we find to be $0.37$ events, in good agreement with the reported $0.36$ events. WE simply assume a background uncertainty of $10\%$ for all components separately.

\begin{figure}[t]
  \centering
  \includegraphics[width=0.99\linewidth]{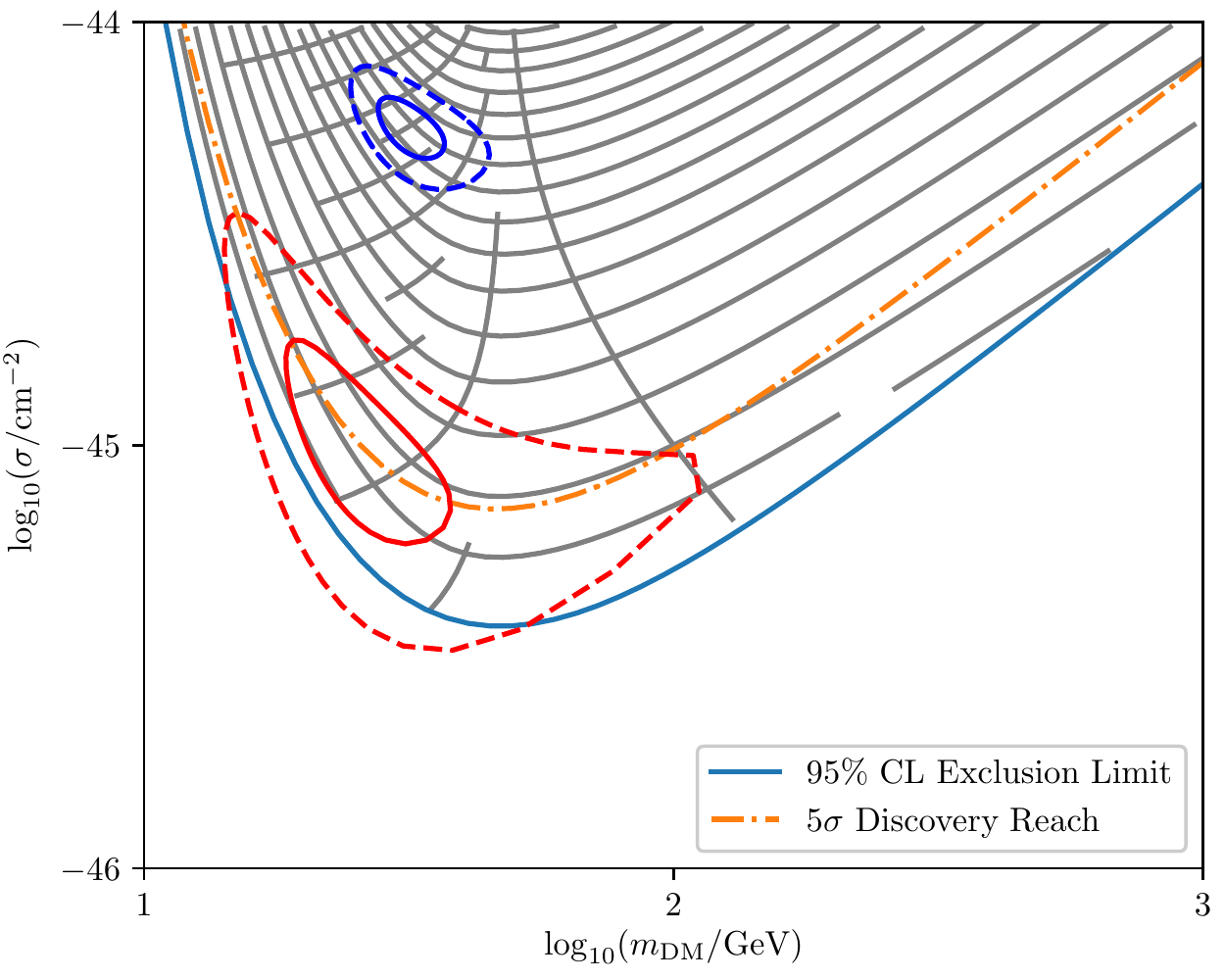}
  \caption{Blue solid line and orange dot-dashed show the $95\%$CL exclusion limit and $5\sigma$ discovery reach respectively. We also plot two examples of $1\sigma$ and $2\sigma$ contours. The grey lines are $1\sigma$ streamline visualisation, as described in the previous sections.}
  \label{fig:XeStreams}
\end{figure}

Again, the implementation of the present example can be found in the
\texttt{jupyter} notebook \texttt{swordfish\_DD.ipynb}.  In
Fig.~\ref{fig:XeStreams}, we show the $95\%$CL projected upper limits, the
$5\sigma$ discovery threshold, example reconstruction contours (68.3\% and
95.4\% CL), as well as the streamline visualization of the Fisher information
metric.  As it is clear from the code, all this is relatively straightforward
to obtain once the instrumental details are coded up.

\section{Conclusions}
\label{sec:Conclusions}

\hyphenation{ana-lyse}

We introduced \swordfish, a set of new statistical tools and their
implementation in a \textsf{Python} package to efficiently forecast and analyse
experimental sensivities in particle and astroparticle physics.  Internally,
\swordfish\ is build on the Fisher information formalism and several new
extensions that were introduced here or recently in
Ref.~\cite{Edwards:2017mnf}.  This allows one to skip the often time-intensive
Monte Carlo step when performing sensitivity forecasts, and to look at
experimental abilities in new and more fine-grained ways.  The tool can be
immediately applied to many practical problems, as we demonstrate with typical
forecasting examples from indirect and direct dark matter searches (the code
examples can be found here at \href{http://www.github.com/cweniger/swordfish}{github.com/cweniger/swordfish}).

The statistical model that we implemented in \swordfish\ is a Poisson
likelihood function, profiled over general Gaussian background perturbations.
Non-linear signal models are supported.  This set-up is generic
enough to capture the relevant aspects of a very large number of experiments
and signal types relevant for particle physics, astroparticle physics and
astronomy (including, for example, collider experiments, direct dark matter
searches, radio telescopes, gamma-ray detectors, neutrino detectors, many
searches for axion-like particles, and even the analysis of gravitational wave
forms).

\medskip

In \swordfish, we implement several new statistical methods and efficient
approximation techniques.
\begin{itemize}
  \item \textit{Equivalent counts method.}  This approximation technique maps
    complex signal and background models onto just two numbers: the
    \textit{equivalent signal} and \textit{equivlanet background} counts.
    These provide information about the number of statistically relevant signal
    and background events, and are used to quickly derive approximate (but accurate)
    \textit{discovery thresholds} and \textit{expected upper limits} on the
    signal flux.
  \item \textit{New visualization of parameter degeneracies.}
    The commonly shown expected confidence regions of benchmark signal models have a
    geometric interpretation as equi-geodesic-distance contours of an
    underlying information metric.  This information metric captures all
    possible benchmark models \textit{at once}.  We introduce an intuitive
    variable-density streamline visualization of this information metric as a
    new way to visualize experimental abilities in a benchmark-free way.
  \item \textit{Fisher information flux.}  In absence of systematic
    uncertainties, the commonly used signal-to-noise ratio is a faithful
    measure to identify the most sensitive region-of-interest in event space.
    We generalize this concept to also account for systematic background
    uncertainties.  The resulting \textit{Fisher information flux} can be used
    for \textit{optimal experimental design} in situations where background
    uncertainties cannot be ignored.
  \item \textit{Euclideanized signal method.}  It is often of interest to
    quantify by how much future experiments will be able to discriminate
    different parts of a signal parameter space.  If the parameter space is
    high-dimensional, this might require the pair-wise comparison of more than
    100 million parameter points.  This can be, in principle, achieved with
    modern clustering algorithms that work in Euclidean space.  In \swordfish\
    we implemented a new way to map signals on their Euclidean analog, such
    that the Euclidean distance between two signals approximates (with high
    accuracy) the statistical difference between the model parameter points.
\end{itemize}

We presented short code examples for how all of the above methods can be easily
used for a specific background and signal model.  The approximation techniques
usually work to within $10\%$--$20\%$, and are validated in
Ref.~\cite{Edwards:2017mnf} or in the present work.  We finally, showed two
simple real-physics examples from both indirect and direct dark matter detection to
illustrate the ease of computation and utility of the method.  Firstly, the
sensitivity to TeV scale annihilating WIMP dark matter with a CTA-like
instrument was calculated, following~\cite{Silverwood:2014yza}.  Secondly, we
constructed a simplified version of the Xenon-1T experiment using an
approximate 1-D analysis using the background signals presented in
Ref.~\cite{Aprile:2017iyp}. For both examples we showed our visualisation schemes as a unique way to view
the model parameter space.

\acknowledgements

TE would like to greatly thank Bradley Kavanagh for discussions on the Xenon1T
signal calculation. We are also happy to thank Jan Conrad, Kyle Cranmer, Pat
Scott and Roberto Trotta for useful discussions. This research is funded by NWO
through the Vidi research grant 680-47-532.

\bibliography{swordfish}

\appendix

\section{Definitions and Derivations}
\label{apx:Defs}
Here we gather all the definitions required to understand the \swordfish package. We avoid reproducing all the derivations of Ref.~\cite{Edwards:2017mnf} and refer there for a more complete set of information.

\subsection{Fisher information matrix}
\label{apx:FIM}

In general, the Fisher information matrix is, given some regularity conditions,
defined via the second derivative of the log-likelihood,
\begin{equation}
  \mathcal{I}_{kl}(\vect\eta)  =   -\left\langle\frac{\partial^2
    \ln \mathcal{L}(\mathcal{D}|\vect{\eta})
    }{\partial \eta_k\partial
  \eta_l} \right\rangle_{\mathcal{D}(\vect\eta)}\;,
  \label{eqn:I}
\end{equation}
where the average is taken over data realizations of a model with parameters
$\vect\eta$.  If we apply this definition to the model in
Eq.~\eqref{eqn:model}, with $\vect\eta = (\vect\theta, \vect{\delta B})$, we
obtain the various Fisher matrix components
\begin{equation}
  \mathcal{I}_{\theta_k\theta_l}(\vect\theta) = 
  \sum_i
  \frac{\partial S_i}{\partial \theta_k}
  \frac{E_i}{S_i(\vect\theta) + B_i}
  \frac{\partial S_i}{\partial \theta_l}\;,
\end{equation}
\begin{equation}
  \mathcal{I}_{\delta B_k\delta B_l}(\vect\theta) = 
  \delta_{kl}\frac{E_k}{S_k(\vect\theta) + B_k} + K^{-1}_{kl}\;,
\end{equation}
\begin{equation}
  \mathcal{I}_{\theta_k\delta B_l}(\vect\theta) = 
  \mathcal{I}_{\delta B_l\theta_k}(\vect\theta) = 
  \frac{\partial S_l}{\partial \theta_k}
  \frac{E_l}{S_l(\vect\theta) + B_l}\;.
\end{equation}
Here, we assumed that the mock data is generated for $\vect\eta = (\vect
\theta, 0)$.

The inverse of the Fisher matrix provides information about the covariance of
the parameters.  However, we are here not interested in the covariance of
$\vect{\delta B}$.  It is hence useful to consider some partially inverted, or
\textit{profiled} Fisher information matrix.  More specifically, the above
Fisher information matrix has block form,
\begin{equation}
  \mathcal{I}_{\vect\eta\vect\eta} =
  \begin{pmatrix}
    \mathcal{I}_{\vect\theta\vect\theta} & \mathcal{I}_{\vect\theta\vect{\delta B}} \\
    \mathcal{I}_{\vect{\delta B}\vect{\delta B}} &
    \mathcal{I}_{\vect{\delta B}\vect\theta}
  \end{pmatrix}\;.
  \label{eqn:block}
\end{equation}
The `profiled Fisher information' matrix for parameters $\vect\theta$ only is
then given by
\begin{equation}
  \mathcal{I} = \mathcal{I}_{\vect\theta\vect\theta} -
  \mathcal{I}_{\vect\theta\vect{\delta B}} \, \mathcal{I}_{\vect{\delta
  B}\vect{\delta B}}^{-1} \,\mathcal{I}_{\vect{\delta B}\vect\theta}\;.
  \label{eqn:Imarg}
\end{equation}
One can now show (see Ref.~\cite{Edwards:2017mnf}) that this expression can be
written in the simple form
\begin{equation}
  \mathcal{I}_{lk}(\vect\theta) = 
  \sum_{ij}
  \frac{\partial S_i}{\partial \theta_k}
  D_{ij}^{-1}
  \frac{\partial S_j}{\partial \theta_l}\;,
\end{equation}
where $D$ is given by
\begin{equation}
  D_{ij} = K_{ij} + \delta_{ij} \frac{S_i(\vect\theta) + B_i}{E_i}\;.
\end{equation}
This is the definition of the Fisher information matrix that is used in
\swordfish.

\subsection{Fisher information flux}

A common question in experimental design, or the planning of observation
strategies, is how much additional data in different areas of observational
parameter space would strengthen the constraints on various model parameters.
The naive approach, which is to just consider signal-to-noise ratios, does not
take into account degeneracies with other model components or background
systematics.  The Fisher information matrix provides a useful starting point to
find a more general definition of the SNR that takes these effects into
account.

We define the Fisher information flux as the partial derivative of the
information matrix $\mathcal{I}_{kl}(\vect\theta)$ w.r.t.~exposure $E_m$,
\begin{equation}
  \mathcal{F}_{kl,m} \equiv \frac{\partial \mathcal{I}_{kl}}{\partial E_m}\;.
\end{equation}
For the above model, this gives
\begin{equation}
  \mathcal{F}_{kl,m} = 
  \sum_{ij}
  \frac{\partial S_i}{\partial \theta_k}
  D_{im}^{-1}
  \frac{S_m(\vect\theta)+B_m}{E_m^2}
  D_{mj}^{-1}
  \frac{\partial S_j}{\partial \theta_l}\;.
\end{equation}
Note that in \swordfish, only a one-dimensional linear model is implemented,
although the generalization to multiple dimensions is straightforward (see
  Ref.~\cite{Edwards:2017mnf}).

\subsection{Equivalent counts method}
\label{apx:ECmethod}

The definition of equivalent signal and background events in
Eqs.~\eqref{eqn:si} and~\eqref{eqn:bi} can be derived as follows.  Consider a
one-bin Poisson process
and a simple linear model $\mu(\theta) = \theta S + B$.  The expected number of
signal and background events are given by $s=\theta S$ and $b=B$, respecively.
These are related to the Fisher information matrix via
$\theta^2\mathcal{I}_{\theta\theta}(\theta) = s^2/(s+b)$.  Evaluating this
relation both at $\theta1>0$ and in the limit $\theta\to0$ gives two equations.
These can be inverted to write the number of signal and background events in
terms of the Fisher information matrix.  We find $s =
\theta^2(\mathcal{I}^{-1}_{\theta\theta}(\theta)-\mathcal{I}^{-1}_{\theta\theta}(0))^{-1}$
and $b = \theta^2\mathcal{I}^{-1}_{\theta\theta}(0)
(\mathcal{I}^{-1}_{\theta\theta}(\theta)-\mathcal{I}^{-1}_{\theta\theta}(0))^{-2}$.
The main idea of the equivalent counts method is now to generalize these exact
expressions to arbitrary models of the form Eq.~\eqref{eqn:model}.  Expected
upper limits and discovery thresholds can then be derived using the equivalent
counts in a hypothetical one-bin Poisson process, instead of the full model.
We showed in Ref.~\cite{Edwards:2017mnf} that the equivalent counts method is
rather accurate for a large variety of cases.  In particular, it is by
construction accurate in the deeply Poissonian and Gaussian regimes.  For
intermediate cases, we found that the obtained limits and thresholds agree with
fully coverage-corrected Monte Carlo results to within $10$--$30\%$.  The most
extreme deviations that we could identify occur for projected upper limits, in
scenarios with the majority of the signal events are buried under a much larger
background and a small amount of signal events in a nearly background free
region dominates the signal significant.  In this case, a kind of
`level-splitting' of the otherwise discrete Poisson likelihood occurs that can
be exploited to obtain somewhat stronger limits than what is possible with a
pure Poisson likelihood.  In this specific case our projected upper limits are
a conservative estimate (see Ref.~\cite{Edwards:2017mnf} for some more
details).

\subsection{Information Geometry}
\label{apx:infoGeom}

Confidence regions at $(1-\alpha)$ confidence level (CL) have the property
that they cover in repeated experiments the true but unknown value in
$(1-\alpha)$ of the cases.  They are very commonly derived by studying
profile likelihood ratios (see, e.g., Ref.~\cite{Cowan:2010js}).  Simple
approximations to the expected confidence regions can be found by assuming that
the likelihood function has the parameter dependence of a multi-variate normal
distribution (`Gaussian approximation').  However, the viability of the
Gaussian approximation depends on the model parameterization.  A more accurate
and parametrization invariant estimate of the expected confidence regions is
based on the geodesic equation.

\subsubsection{Gaussian approximation}

If the Gaussian approximation applies, the expectation-value of the
log-likelihood function has the form
\begin{equation}
  -2\left\langle
  \frac{\ln\mathcal{L}(\mathcal{D}|\vect\theta')}
  {\ln\mathcal{L}(\mathcal{D}|\vect\theta)}
  \right\rangle_{\mathcal{D}(\vect\theta)}
  = 
    \sum_{ij}
    (\theta_i-\theta'_i)
    \mathcal{I}_{ij}(\vect{
    \theta})
  (\theta_j-\theta'_j)
  \equiv 
  \chi^2\;.
\end{equation}
The expected confidence region at $(1-\alpha)$CL, provided that
$\vect\theta$ has been measured, can be obtained by identifying the parameter
region where 
\begin{equation}
  \chi^2(\vect\theta') \leq Y_k^2\;,
\end{equation}
where $Y_k$ is defined in Eq.~\ref{eqn:Ychi2} and $k$ corresponds to the dimensionality of the parameter space
$\vect\theta$.

\subsubsection{Geodesic approximation}
\label{subsec:Geodesic}

The above construction of confidence contours has a geometric interpretation.
The Fisher information matrix induces a \textit{metric} on the parameter space
of $\vect\theta$.  In fact, a central property of the Fisher information matrix
is that it transforms like a metric under reparametrization of the model.  This
follows directly from the definition in Eq.~\eqref{eqn:I}.  The region defined
above corresponds then to an equal-geodesic-distance contour with distance $Y$
from the point $\vect \theta$ (provided that the Fisher information matrix does
not vary significantly within the region of interest).
A \textit{parametrization-independent definition} of confidence regions can be
found by constructing equal-geodesic-distance contours by solving the geodesic
equation,
\begin{equation}
  \frac{d^2\theta_i}{ds^2} +
  \frac12\mathcal{I}_{ij}^{-1}
  \left(
    \frac{\partial \mathcal{I}_{lj}}{\partial \theta_k}
    +\frac{\partial \mathcal{I}_{kj}}{\partial \theta_l}
    -\frac{\partial \mathcal{I}_{kl}}{\partial \theta_j}
  \right)\frac{d\theta_k}{ds}\frac{d\theta_l}{ds}=0\;.
\end{equation}
Although this \textit{geodesic approximation} to confidence contours is still
not exact, we find that they in general agree very well with results from
traditional likelihood-ratio explorations, both in the statistics and
systematic limited regime.

\subsection{Euclideanized signal}

The entire point of the `euclideanized signals' that we introduce in this work
to enable the rapid calculation of the expected TS value in Eq.~\eqref{eqn:xy}
for a \textit{very} large number of signal combinations (\text{e.g.},
$\sim10^{16}$ in the case of a Bayesian scan with 100 million points).  This
would be possible if the most efficient clustering algorithms, which can handle
billions of points, can be used.  These, however, happen to work in Euclidean
space.  The goal is hence to map signals $\vect S$ onto vectors $\vect x$ such
that the TS value is approximately given by the L2-norm in Euclidean space,
$\text{TS}\simeq \lVert \vect x_1 - \vect x_2 \rVert^2$.

We start by approximating the TS-value in Eq.~\eqref{eqn:xy} by using the
Fisher information matrix evaluated at the mean signal $\bar {\vect S} =
\frac12 (\vect S_1+\vect S_2)$,
\begin{align}
  TS 
  \simeq  \Delta \vect S^T D_{\bar{\vect S}}^{-1}  \Delta \vect S\;,
  \label{eqn:xy}
\end{align}
where $\Delta \vect S \equiv \vect S_1 - \vect S_2$ denotes the signal
difference, and $D_{\bar{\vect S}}$ corresponds to Eq.~\eqref{eqn:D} with
$S_i\to \bar S_i$.  If we define now the vector
\begin{equation}
  x'_i \equiv \sum_j (D^{-1/2})_{ij} S_j E_j\;,
\end{equation}
it is straightfoward to show that it satisfies
\begin{equation}
  TS \simeq \lVert \vect x'_1 - \vect x'_2 \rVert^2 \quad\text{if}\quad
  D_{\vect S_1} \approx D_{\vect S_2}\;.
\end{equation}
However, in the case where the shot noise of the signal has a significant
impact on the total background uncertainties, the above relationship will break
down.  This becomes actually a large effect in the strong-signal limit.  In
fact, to second order in $\Delta S_i^2$, we find that
\begin{equation}
  \lVert \vect x'_1 - \vect x'_2 \rVert^2 
  \approx  \frac14 \sum_i \frac{\Delta S_i^2}{\bar S_i}
  \quad\text{if}\quad \bar S_i \gg B_i + E_i K_{ii}\;.
\end{equation}
Hence, we would in this limit underestimate the discrimination power of an
instrument systematically by a factor of two.  In order to avoid this problem,
we multiply $\vect x'$ with some \textit{fudge factor} that equals one for
negligible signals, and equals two when the signal dominates.  We will show below that
this proceedure leads to satisfactory results.

\medskip

The above discussion motivates the definition of the `Euclideanized signal',
\begin{equation}
  x_i \equiv \left(\sum_j (D^{-1/2})_{ij} S_j E_j\right)
  \left( 1+ \frac{R\cdot S_i}{R\cdot S_i + B_i + K_{ii}E_i} \right)\;,
  \label{eqn:eucsig}
\end{equation}
where we take the weight $R=0.1$, which we found to lead to the best results.
It has the proprety that the TS-value corresponding to two signals can now be
written as Euclidean distance between their euclideanized signal vectors, as shown in Eq.~\eqref{eqn:xy}.

\medskip

In order to test the accuracy of the above proceedure, we randomly generated a
large number of models with $n_\text{b}=10$, using randomized signals $\vect S$
and randomized covariance matrices $K$.  Without loss of generality, the
background and the exposure are kept flat.  In Fig.~\ref{fig:validation2}, we
compare the TS values obtained from the profile log-likelihood ratio in
Eq.~\eqref{eqn:xy} with the one obtained from the euclideanized signals, for
varying degrees of exposure and magnitude of the covariance matrix. 

\section{Validation of approximation methods}
\label{apx:validation}

First, we motivate heuristically the analytical form of the fudge factor that
we used in Eq.~\eqref{eqn:eucsig}.  To this end, we consider the approximate
relation
\begin{equation}
  \chi^2_G \equiv  \frac{(s_1-s_2)^2}{b+\bar s} \approx 
  \chi^2_E \equiv (x_1-x_2)^2\;,
  \label{eqn:eutest}
\end{equation}
where we defined the mean signal $\bar s \equiv \frac12(s_1 + s_2)$, and the Euclideanized signal
\begin{equation}
  x_i = \frac{s_i}{\sqrt{s_i+b}} \cdot \left( 1 + \frac{R\cdot s_i}{R\cdot s_i
  + b} \right)
\end{equation}
As above, we set the rescaling parameter $R=0.1$, for which we find the best
performance.

\begin{figure}
  \centering
  \includegraphics[width=\linewidth]{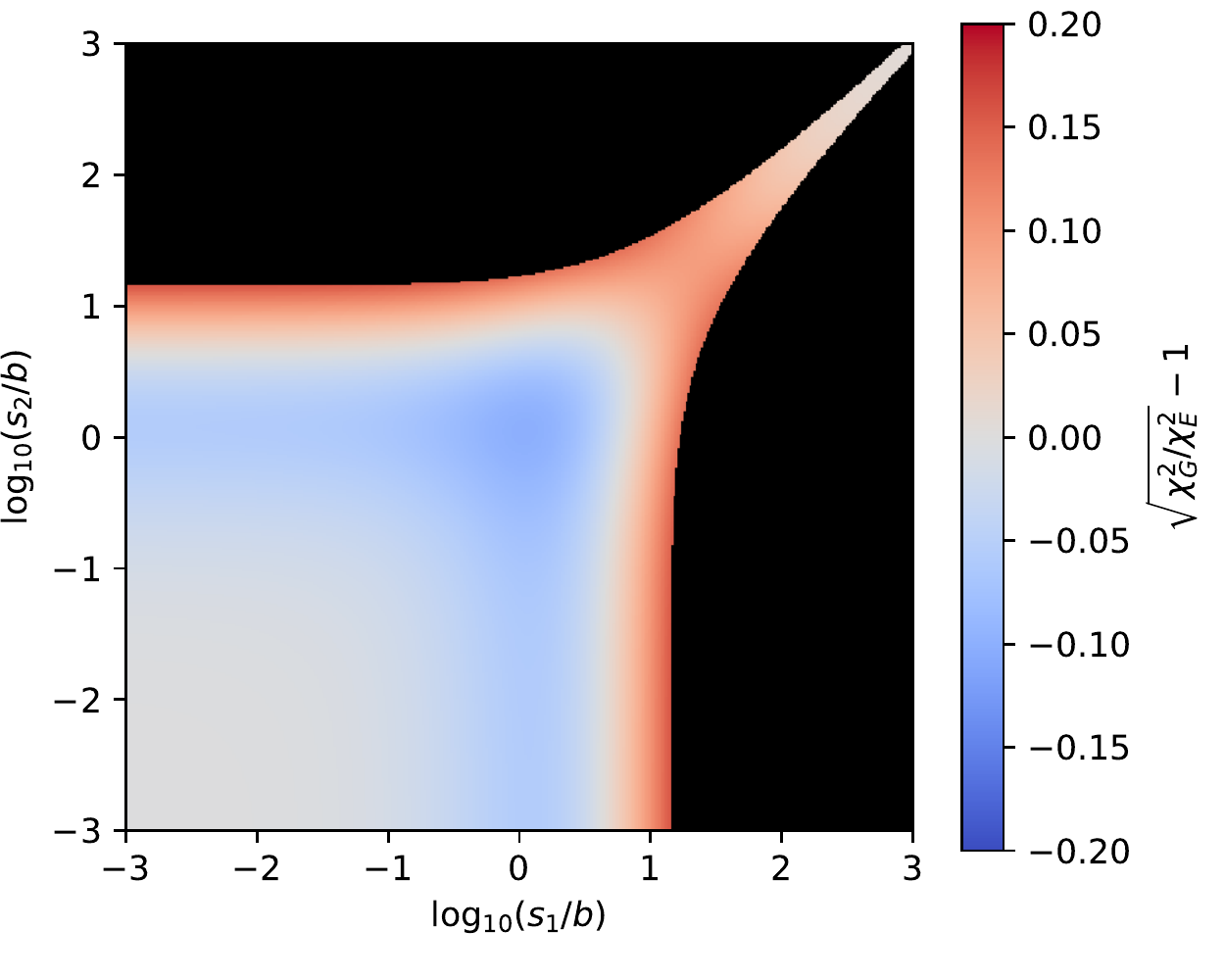}
  \caption{Ratio between $\sqrt{\chi^2_G}$ and $\sqrt{\chi_E^2}$ in
  Eq.~\eqref{eqn:eutest}, as function of $s_1/b$ and $s_2/b$.  The unmasked
  area corresponds to $\chi^2_G < 25$, which excludes signal differences larger than $\sim5\sigma$.  Within the unmasked region, the differences between $\sqrt{\chi^2_G}$ and $\sqrt{\chi^2_E}$ are smaller than 16\%.}
  \label{fig:validation1}
\end{figure}

The degree to which the approximation in Eq.~\eqref{eqn:eutest} is shown in
Fig.~\ref{fig:validation1}, as function of the signals $s_1$ and $s_2$.  Only
the region that corresponds to a signal difference of $<5\sigma$ is shown.  In
this region, the approximation provides a relative agreement to within $16\%$.
The approximation works somewhat worse if the difference between the signals
becomes larger, which is however not of much relevance for the intendet
applications of our method.

\medskip

In order to validate the Euclideanized signal method, and to estimate the
expected approximation errors, we consider a large number of random models.
For the purpose of illustration in this paper, we kept the models simple.  They
consists of $\nbins=3$ bins (we find similar results also for a much larger
number of bins), with a background set to $\vect B^T = (1, 1, 1)$, a signal set
to $\vect S = \theta \vect R$, where $\vect R$ is a vector of random numbers in
the range $[0, 1]$, and $K= k^2 L^T\cdot L$ where $L$ is a random $3\times 3$
matrix with entries in the range $[-1, 1]$.  We assume a flat exposure given by
$E$.  We consider three benchmark scenarios. First, a \textit{signal limited}
case, with $E=10^{-2}$, $k = 0$, $\theta = 10^{3.5}$.  Second, a
\textit{systematics limited} case, with $E= 10^6$, $k = 1$, $\theta = 1$.
Third, a \textit{background limited} case that is still close to the Poissonian
regime, with $E= 10^2$, $k = 0$, $\theta=1$.

\begin{figure}
  \centering
  \includegraphics[width=\linewidth]{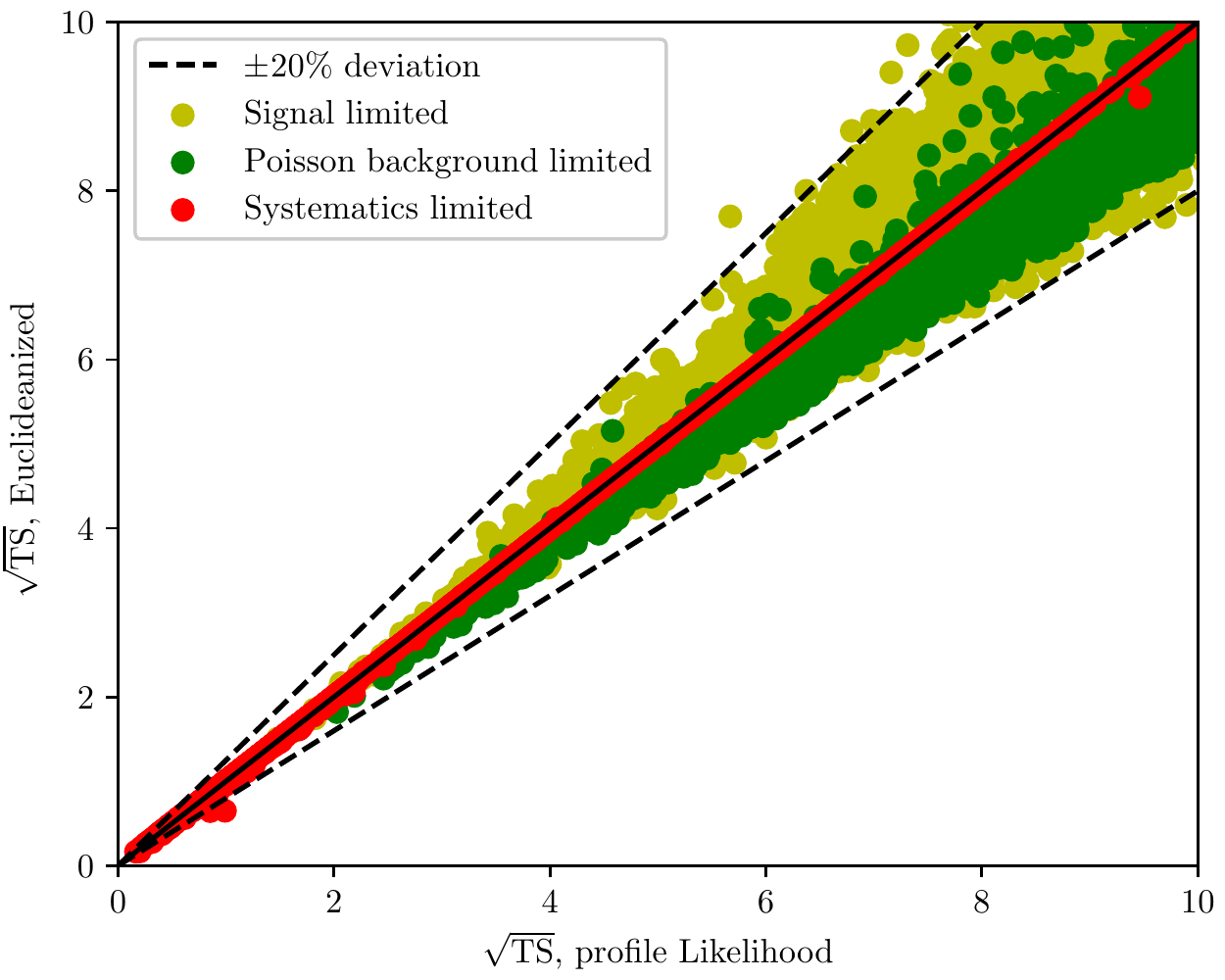}
  \caption{Comparison between exact $\sqrt{\rm TS}$ value derived from the log
  profile-likelihood, and the approximate $\sqrt{\rm TS}$ value derived from
  the Euclideanized signal method.  The different colors correspond to the
  signal, the systematics and the (Poisson) background limited regimes (see text
  for details).  As indicated by the dashed lines, the devitations are not larger
  than $20\%$ for all random models.}
  \label{fig:validation2}
\end{figure}

In Fig.~\ref{fig:validation2} we co-eps-converted-to.pdfmpare the TS-value derived via the profile
log-likelihood, Eq.~\eqref{eqn:TScorrect}, with the TS-value derived from the
Euclideanized signal method, Eq.~\eqref{eqn:TSeuc}.  We find that the
deviations are largest in the signal-limited case, and up to $\pm20\%$ on
$\sqrt{TS}$.  In the systematics limited regime, the deviations are (as
expected) much smaller.  Interestingly, in the background limited case, we find
for the given benchmark point that there is a $\sim 10\%$ bias towards smaller
TS-values.  This bias disappears if we either increase the exposure (and hence
the problem becomes more Gaussian), or if we decrease the exposure and the
problem becomes signal dominated.  We conclude that \textit{the Euclideanized
signal method provides fast estimates for $\sqrt{\rm TS}$ that are correct to
within $20\%$.}

\end{document}